\documentclass[aps,a4paper,twocolumn,english, superscriptaddress,eqsecnum,showpacs,longbibliography]{revtex4-1}
\usepackage{babel}
\usepackage{xcolor}
\usepackage{graphicx}
\usepackage{amsmath}
\usepackage{amssymb}
\usepackage{wasysym}
\usepackage{hhline}

\newcommand{\bc}{\begin{center}}
\newcommand{\ec}{\end{center}}
\newcommand{\be}{\begin{equation}}
\newcommand{\ee}{\end{equation}}
\newcommand{\bea}{\begin{eqnarray}}
\newcommand{\eea}{\end{eqnarray}}

\definecolor{darkblue}{rgb}{0.1,0.2,0.6}
\definecolor{darkred}{rgb}{0.8,0.1,0.2}
\usepackage[colorlinks,
            citecolor=darkblue,
            linkcolor=darkred,
            urlcolor=darkblue]{hyperref}

\newcommand{\nuexp}{0.75(11)}
\newcommand{\betaexp}{1.08(20)}
\newcommand{\hcval}{12.30(2)}
\newcommand{\etaexp}{-1.12(10)}

\usepackage[all]{hypcap} 

\makeatletter
\adddialect\l@English\l@english
\makeatother

\graphicspath{{./images/}}

\begin{document}
\title{Competing Bose-Glass physics with  disorder-induced Bose-Einstein condensation in the doped $S=1$ antiferromagnet
Ni(Cl$_{1-x}$Br$_x$)$_2$-4SC(NH$_2$)$_2$ at high magnetic fields}
\author{Maxime Dupont}
\email{maxime.dupont@irsamc.ups-tlse.fr}
\affiliation{Laboratoire de Physique Th\'eorique, IRSAMC, Universit\'e de Toulouse, CNRS, UPS, France}
\affiliation{Department of Physics, Boston University, 590 Commonwealth Avenue, Boston, Massachusetts 02215, USA}
\author{Sylvain Capponi}
\email{sylvain.capponi@irsamc.ups-tlse.fr}
\affiliation{Laboratoire de Physique Th\'eorique, IRSAMC, Universit\'e de Toulouse, CNRS, UPS, France}
\affiliation{Department of Physics, Boston University, 590 Commonwealth Avenue, Boston, Massachusetts 02215, USA}
\author{Mladen Horvati\'c}
\email{mladen.horvatic@lncmi.cnrs.fr}
\affiliation{Laboratoire National des Champs Magn\'etiques Intenses, LNCMI-CNRS (UPR3228), \\ EMFL, UGA, UPS, and INSA, Bo\^{i}te Postale 166, 38042, Grenoble Cedex 9, France}
\author{Nicolas Laflorencie}
\email{nicolas.laflorencie@irsamc.ups-tlse.fr}
\affiliation{Laboratoire de Physique Th\'eorique, IRSAMC, Universit\'e de Toulouse, CNRS, UPS, France}
\date{\today}

\begin{abstract}
    We study the interplay between disorder and interactions for emergent bosonic degrees of freedom induced by an external magnetic field in the Br-doped spin-gapped antiferromagnetic material Ni(Cl$_{1-x}$Br$_x$)$_2$-4SC(NH$_2$)$_2$ (DTN$X$). Building on nuclear magnetic resonance experiments at high magnetic field [A. Orlova {\it{et al.}}, \href{http://journals.aps.org/prl/abstract/10.1103/PhysRevLett.118.067203}{Phys. Rev. Lett. {\bf 118}, 067203 (2017)}], we describe the localization of isolated impurity states, providing a realistic theoretical modeling for DTN$X$. Going beyond single impurity localization we use quantum Monte Carlo simulations to explore many-body effects from which pairwise effective interactions lead to a (impurity-induced) Bose-Einstein condensation (BEC) revival [M. Dupont, S. Capponi, and N. Laflorencie, \href{http://journals.aps.org/prl/abstract/10.1103/PhysRevLett.118.067204}{Phys. Rev. Lett. {\bf 118}, 067204 (2017)}].  We further address the question of the existence of a many-body localized Bose-glass (BG) phase in DTN$X$, which is found to compete with a series of a new kind of BEC regimes made out of the multi-impurity states. The global magnetic field--temperature phase diagram of DTN$X$ reveals a very rich structure for low impurity concentration, with consecutive disorder-induced BEC mini-domes separated by intervening many-body localized BG regimes. Upon increasing the impurity level, multiple mini-BEC phases start to overlap, while  intermediate BG regions vanish.
\end{abstract}

\maketitle

\section{Introduction}\label{sec:introduction}

In condensed matter physics, simple but faithful theoretical models are derived from relevant degrees of freedom and interactions in realistic materials. They aim to capture and describe low-energy properties, including, for instance, exotic phases or phase transitions~\cite{sachdev2001,dagotto2005}. Unlike classical phase transitions driven by thermal fluctuations, quantum phase transitions (QPT)~\cite{coleman2005,sachdev2011} happen at exactly zero temperature and are driven by external parameters such as pressure, magnetic field or disorder. In this paper we address the antiferromagnetic insulator compound NiCl$_2$-4SC(NH$_2$)$_2$ (DichlorotetrakisThioureaNickel, or DTN for short), which is well-known to display a Bose-Einstein condensation (BEC) --- corresponding to an antiferromagnetic (AF) ordered phase~\cite{giamarchi2008} --- upon applying a sufficiently strong external magnetic field. When doping with Br impurities, Ni(Cl$_{1-x}$Br$_x$)$_2$-4SC(NH$_2$)$_2$ (DTN$X$) is known to display fascinating properties~\cite{yu2012nature}: it was reported as one of the first realizations of the many-body localized Bose-glass (BG) phase in a quantum magnet, providing a possibility for experimental investigations of the critical properties of the BEC--BG transition.
This gave rise to a thorough discussion about the experimental and numerical values of the critical exponents for such a transition~\cite{yu2012,yao2014,yu2014,Syromyatnikov2017} as compared to the Fisher's theory~\cite{fisher1989}. We start this work by quickly reviewing these two unconventional phases, namely the BEC and the BG, focusing on their realization in quantum magnets and more specifically in DTN($X$).

\subsection{Magnetic field-induced Bose-Einstein condensation in quantum antiferromagnets}

The Bose-Einstein condensation was first introduced in the context of bosons and superfluid ${}^4$He~\cite{london1938,pitaevskii2003}, where a macroscopic number of particles occupies the lowest-energy state below a critical temperature $T_c$. It was then realized through the mapping between spins and bosons~\cite{matsubara1956} that BEC can be produced in many quantum antiferromagnets under magnetic field~\cite{giamarchi1999,nikuni2000,rice2002,giamarchi2008}, see Ref.~\onlinecite{zapf2014} for a complete review. This can be understood as the condensation of spin excitations, leading to a spontaneous breaking of the continuous $\mathrm{U}(1)$ symmetry below $T_c$, and to Nambu-Goldstone modes~\cite{nambu1960,goldstone1962} with a linear dispersion above the BEC ground state (GS). In terms of the underlying magnetic degrees of freedom, the BEC is equivalent to transverse XY-order. This enthralling property was explored both theoretically and experimentally for the coupled ladders compounds Cu$_2$(C$_5$H$_{12}$N$_2$)$_2$Cl$_4$~\cite{giamarchi1999} and CuBr$_4$(C$_5$H$_{12}$N)$_2$~\cite{klanjsek2008,bouillot2011} or the dimer systems TlCuCl$_3$~\cite{nikuni2000,ruegg2003} and BaCuSi$_2$O$_6$~\cite{jaime2004,sebastian2006,ruegg2007}.

\begin{figure}[ht]
    \includegraphics[width=\columnwidth,clip,angle=0]{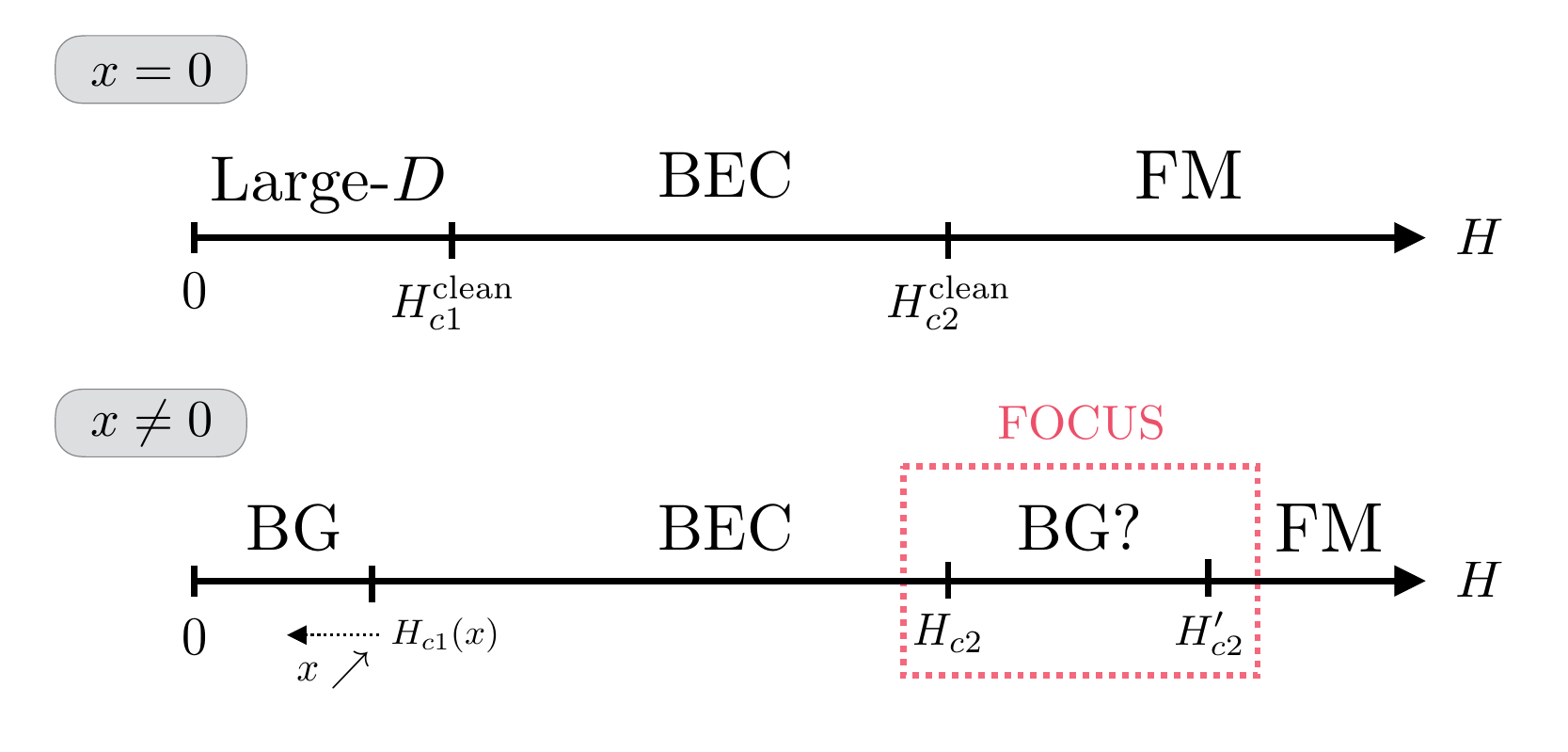}
    \caption{Zero temperature phase diagram of DTN ($x=0$) and DTN$X$ ($x\neq 0$) as a function of the magnetic field $H$. In the clean case, the intermediate long-range ordered phase (BEC) is bounded by two critical fields $H_{c1}^\mathrm{clean}$ and $H_{c2}^\mathrm{clean}$. Below $H_{c1}^\mathrm{clean}$, DTN is in a large-$D$ phase due to strong single-ion anisotropy and above $H_{c2}^\mathrm{clean}$, the material is a fully polarized ferromagnet (FM). The disorder $x\ne 0$ expands the DTN$X$ phase diagram with new BG phases. However, the high magnetic field BG phase was recently found to be undermined by long-range ordering induced by the disorder~\cite{orlova2017,dupont2017}. In this work, we focus on this putative BG at high magnetic field (red square). Note also that the first critical field $H_{c1}$ is renormalized downwards by the doping.}
    \label{fig:original_phase_diagram}
\end{figure}

Another example of such a material is the weakly coupled spin-one chains compound DTN~\cite{filho2004} which features three different regimes at low temperature upon applying an external magnetic field as outlined in Fig.~\ref{fig:original_phase_diagram} ($x=0$). From zero field to $H_{c1}^\mathrm{clean}\simeq 2.1\;\mathrm{T}$, the material is in a spin-gapped phase, so-called large-$D$ phase~\cite{botet1983,schulz1986,chen2003,hu2011} due to strong uniaxial single-ion anisotropy ($D$). At higher field and for temperatures below $T_c\sim 1$~K, DTN is magnetically ordered~\cite{zapf2006,zvyagin2007,zvyagin2008,wulf2015,blinder2017}, while above $H_{c2}^\mathrm{clean}=12.3\;\mathrm{T}$, the material is a trivial spin-gapped ferromagnet (FM).

\subsection{Bose-glass physics}

Disorder is intrinsically present in any realistic system, and may play a major role in some observed phenomena such as Anderson localization for non-interacting particles~\cite{anderson1958,evers2008,lagendijk2009}. The effect of interactions and its interplay with disorder in bosonic systems received a great deal of attention since the experiments on superfluid Helium in porous media~\cite{reppy1992}. Subsequent theoretical studies revealed a new many-body localized phase of matter at zero temperature: the Bose-glass state~\cite{giamarchi1988,fisher1989}: an inhomogeneous gapless compressible fluid with short-ranged (exponentially suppressed) correlations. The dimensionality is of great importance; in one dimension (${\rm D}=1$) disorder is a relevant perturbation in most of the cases~\cite{giamarchi2004}, while in D~=~2 a finite disorder strength is required~\cite{alvarez2015,ng2015} to destroy the zero-temperature superfluid condensate, leading to the BG state. For D~=~3 one needs stronger randomness to eventually localize the bosons~\cite{gurarie2009}. The BG phase was observed in disordered superconducting thin amorphous $\mathrm{InO}$ films where superconductivity is destroyed by the localization of the Cooper pairs~\cite{sacepe2011}, or in trapped cold atoms setups~\cite{derrico2014}.

Quantum magnets subject to disorder have shown a wide range of interesting phenomena: from the random singlet phase~\cite{fisher1994,shiroka2011,thede2012} to the order-by-disorder mechanism induced by the impurities~\cite{azuma1997,uchiyama1999,bobroff2009} as well as the BG phase which was theoretically investigated~\cite{nohadani2005,hida2006,roscilde2006} and reported in the (CH$_3$)CHNH$_3$(Cu$_x$Cl$_{1-x}$)$_3$ and Tl$_{1-x}$K$_x$CuCl$_3$ compounds~\cite{hong2010tao,yamada2011}, see Ref.~\onlinecite{zheludev2013} for a recent review. The Br-doped version of the DTN compound, DTN$X$, was recently proposed to be an exceptionally convenient archetype material presenting a BG phase \cite{yu2010,yu2012nature,yu2012}. Both the BEC and the polarized phase are robust to disorder and subsist in the doped DTN$X$ compound, although their critical fields $H_{c1}$ and $H_{c2}'$ are shifted. In addition, new BG regimes are predicted to (i) substitute the gapped large-$D$ regime at low field and (ii) to intervene between the BEC and the polarized phase between $H_{c2}$ and $H_{c2}'$ (see Fig.~\ref{fig:original_phase_diagram} at $x\ne 0$). The BG phase in DTN$X$ can be pictured and defined as follow: coexisting with a gapped background, localized magnetic states occur in the vicinity of impurities and display a finite local susceptibility. These localized degrees of freedom are spatially separated with exponentially decaying correlations which prevent any long-range ordering. Until recent experimental and theoretical works~\cite{orlova2017,dupont2017}, it was proposed~\cite{yu2012nature} that the BG phase at high magnetic field is uninterrupted between the BEC and FM regimes, from $H_{c2}$ to $H'_{c2}$, as shown in Fig.~\ref{fig:original_phase_diagram} for $x\neq 0$. Instead, it turns out that the impurity degrees of freedom display a striking ``many-body delocalization'' with a resurgence of a global phase coherence, leading to disorder-induced long-range order (LRO)~\cite{nohadani2005,hida2006,dupont2017}.

\subsection{Main results and structure of the paper}

Recent nuclear magnetic  resonance (NMR) experiments performed on DTN$X$ at high magnetic field, in the putative BG regime (Fig.~\ref{fig:original_phase_diagram}, $x\neq 0$), have revealed a level-crossing of the impurity states at a magnetic field $H^*= 13.6\;\mathrm{T}$ \cite{orlova2017}. These impurity states are found to be exponentially localized, with very short localization lengths, $\xi_\parallel\simeq 0.48$ and $\xi_\perp\simeq 0.17$ in units of lattice spacings. This local characterization allows (i) to determine the microscopic parameters of DTN$X$ and (ii) to extract the effective \emph{unfrustrated} pairwise interaction between impurities, which eventually leads to a long-range ordering. Although this interaction is exponentially suppressed with the distance $|\mathbf{r}|$ between two localized states, $\propto \exp\left(-|\mathbf{r}|/2\xi_{\parallel,\perp}\right)$, it was numerically shown in Ref.~\onlinecite{dupont2017} that at low temperature appears an inhomogeneous BEC$^*$ regime in a field range around $H^*$, in analogy with disorder-induced ordering mechanism of the order-from-disorder type \cite{villain1980,shender1991}.

\begin{figure}[ht]
    \includegraphics[width=\columnwidth,clip,angle=0]{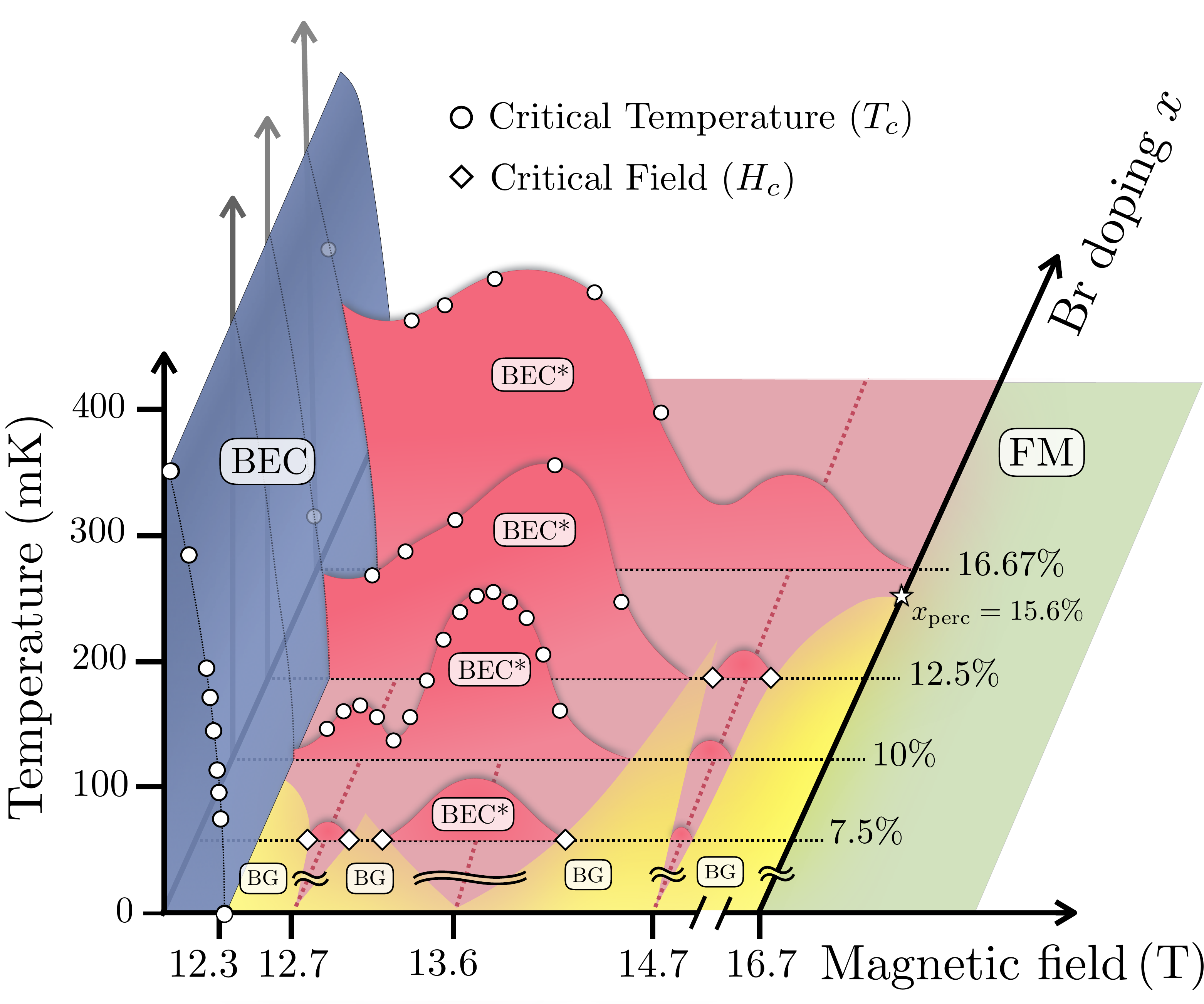}
    \caption{Global Magnetic field -- temperature phase diagram for Ni(Cl$_{1-x}$Br$_x$)$_2$-4SC(NH$_2$)$_2$ (DTN$X$) based on numerical (QMC) results (circles and diamonds), displayed for varying Br doping $x$. For small finite doping $x$, above the clean BEC phase (blue dome) at $H>12.3\;\mathrm{T}$, a succession of impurity-induced BEC$^*$ phases (pink domes) is stabilized together with intervening localized Bose-glass (BG) regimes (yellow regions), before getting into the fully polarized ferromagnet (FM, green region)~\cite{footnote_figure2}. Such a localization-delocalization series is expected to disappear for increasing doping $x$, to eventually form a unique impurity induced BEC$^*$ regime, overlapping with the principal BEC dome.  Above the 3D percolation threshold $x_{\rm perc}=15.6\%$, the system is expected to be ordered at all field values up to the full polarization. This global diagram summarizes the results presented in this paper.}
    \label{fig:phase_diagram_3d}
\end{figure}

This paper is constructed as follows. In section~\ref{sec:theo_model}, we first present a short experimental overview of the material and build a microscopic model for DTN$X$  based on local NMR measurements at high magnetic field, mostly relying on single impurity physics. Section~\ref{sec:manybody_impurity} introduces the building blocks for the impurity-induced long-range ordering within the high magnetic field BG phase, namely the computation of the effective pairwise interaction between the magnetic impurities. We also discuss the many-impurity effects and their experimental evidences in DTN$X$. Then, using large scale numerical simulation based on the quantum Monte Carlo algorithm, we reveal in section~\ref{sec:DTNX_physics} long-range ordering of the impurity degrees of freedom at concentrations and temperatures that should be accessible to experiments. In Sec.~\ref{sec:bose_glass} we show how, upon decreasing the Br-doping concentration, consecutive disorder-induced BEC mini-domes are separated by intervening BG regimes (Fig.~\ref{fig:phase_diagram_3d}). We thus unveil the amazing richness of the high magnetic field phase diagram of DTN$X$, which is shown in Fig.~\ref{fig:phase_diagram_3d} in the three dimensional representation: magnetic field -- temperature -- Br concentration ($H$--$T$--$x$). Finally, Sec.~\ref{sec:conclusion} presents concluding remarks.

\section{Microscopic modeling of DTN$X$}\label{sec:theo_model}

The DTN material is a three-dimensional ($3$D) antiferromagnet consisting of weakly coupled chains of $S=1$ spins, borne by Ni$^{++}$ ions, subject to a strong single-ion anisotropy. The potential interest of this system, presenting at low temperature a magnetic-field-induced, 3D-ordered, canted phase, was realized already in 1981 \cite{paduan1981}, but DTN became a topical system only after this type of phase was recognized to be a convenient representative of the BEC \cite{giamarchi1999,nikuni2000}, and the upper critical field of the BEC phase in DTN is found to be experimentally well accessible, $H_{c2}^\mathrm{clean}=$~12.3~T \cite{filho2004}. Since then, it became one of the most studied archetypal materials for the BEC-type spin systems \cite{zapf2014}. For the purpose of this article we will use the first precise set of the exchange couplings determined for DTN in Ref.~\cite{zvyagin2007} using the BEC phase boundary, the magnetization and the ESR data. This set was further refined by the high-field neutron results \cite{tsyrulin2013} to take into account the frustrated coupling between the two tetragonal subsystems of the DTN's \emph{body-centred} tetragonal lattice. The frustration makes the effects of this coupling negligible, as shown by the numerical analysis of the order parameter in the BEC phase determined by NMR \cite{blinder2017}. To describe pure and doped DTN we will thus use the following model for $S=1$ spins on a simple tetragonal lattice:
\bea
    {\cal{H}}&=&\sum_i\Bigl[\sum_nJ_{i,n}{\boldsymbol{S}}_{i,n}\cdot {\boldsymbol{S}}_{i+1,n} +J_\perp\sum_{\langle n\,m\rangle}{\boldsymbol{S}}_{i,n}\cdot {\boldsymbol{S}}_{i,m}\nonumber\\
    &+&\sum_n D_{i,n}\left(S_{i,n}^{z}\right)^2-g\mu_B H S_{i,n}^z\Bigr],
    \label{eq:DTNX}
\eea
where for pure DTN the AF exchange along the chain direction is $J_{i,n}=J=2.2\;\mathrm{K}$, the single-ion anisotropy is $D_{i,n}=D=8.9\;\mathrm{K}$, and the chains are coupled by the interchain coupling between the nearest-neigbor sites (denoted by $\langle n\,m\rangle$) $J_\perp=0.18\;\mathrm{K}$. $H$ is an external magnetic field applied along the single-ion anisotropy axis $z$, thus preserving the $\mathrm{U}(1)$ symmetry. We use $g=2.31$ for the gyromagnetic factor, such that in the absence of chemical disorder, the clean upper critical field $H^\mathrm{clean}_{c2}=(D+4J+8J_\perp)/g\mu_B=12.3\;\mathrm{T}$, as pictured in Fig.~\ref{fig:original_phase_diagram} ($x=0$).

\begin{figure}[b]
    \includegraphics[width=0.8\columnwidth,clip,angle=0]{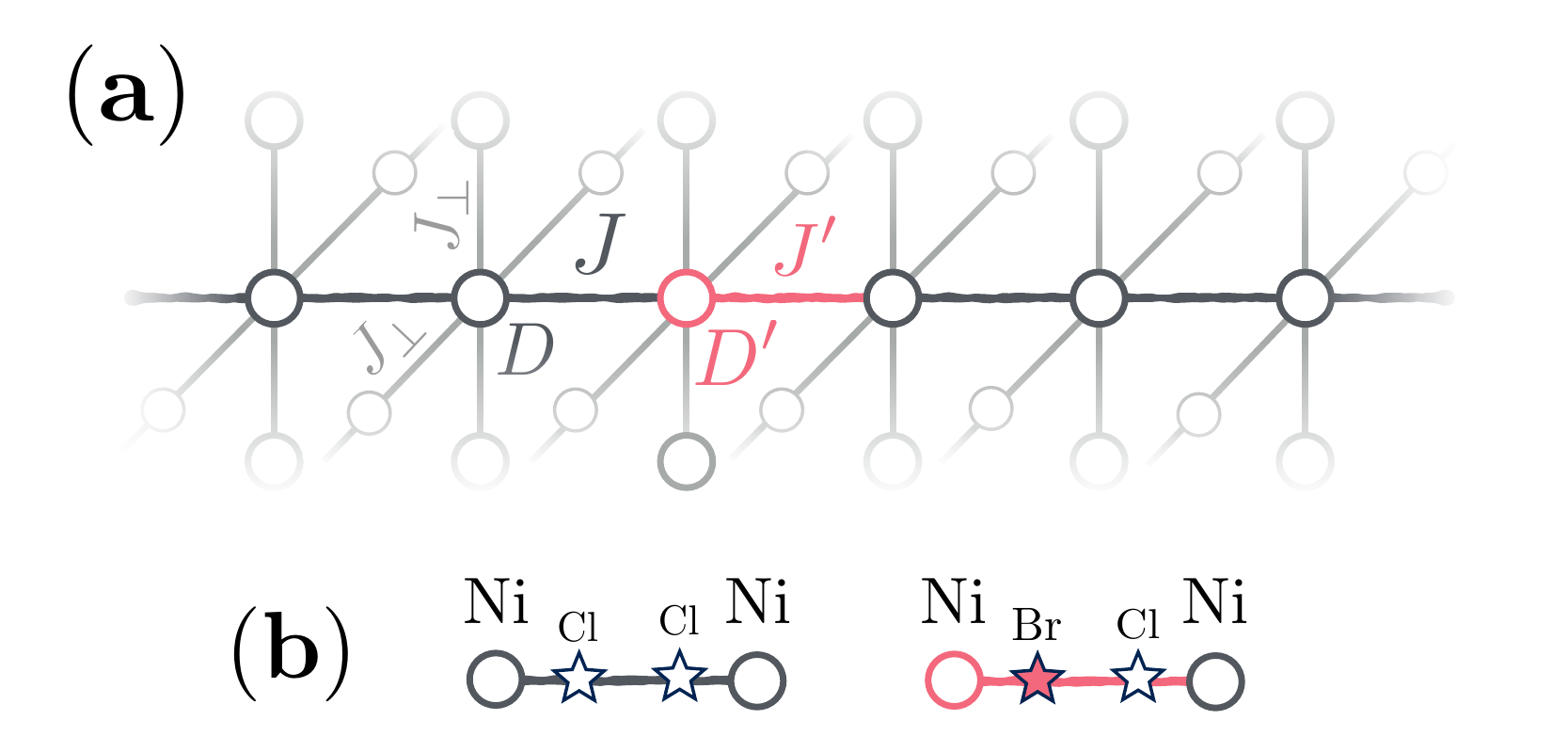}
    \caption{({\bf a}) Sketch representation of the relevant $3$D structure for DTN$X$ model. On the chains, the clean sites (single ion anisotropy $D$) with first-neighbor interaction ($J$) are in grey. The doped ones (single ion anisotropy $D'$) are in pink with the modified interaction ($J'$) in pink as well. The three dimensional coupling between the chains $J_\perp$ is not affected by the doping. For readability, only one thick line representing the main chain is displayed. ({\bf b}) Two types of $S=1$ dimers: clean Cl$-$Cl (left hand side) and doped Br$-$Cl (right hand side), with Br preferentially positionned on the left (see supplementary material of \cite{yu2012nature}).}
    \label{fig:dtnx_sketch}
\end{figure}

In the doped DNT$X$ compound, as shown in Fig.~\ref{fig:dtnx_sketch}(b), one of the two Cl$^-$ ions in the intrachain $J$ coupling bond may be substituted by the doped Br$^-$ ``impurity'', introducing thereby a disorder in the system. Based on the macroscopic experimental data (magnetization, susceptibility and specific heat) and global modeling of the system by quantum Monte Carlo (QMC) simulation, DTN$X$ was proposed to be a model system for the investigation of the BG phase \cite{yu2010,yu2012nature}. The system is modeled assuming that the doping introduces only local perturbations: each Br impurity modifies only the exchange coupling value of the affected bond to $J'=2.42J$ and the single-ion anisotropy of the closest Ni ion to $D'=0.36D$, without affecting any other bond or anisotropy value [Fig.~\ref{fig:dtnx_sketch}(b)]. In comparison to the initially proposed $J'$ and $D'$ values \cite{yu2012nature}, the values given here are refined by combining the recent NMR measurements and theoretical work \cite{orlova2017,dupont2017}, which is explained in this section.

As regards other experimental investigations of the DTN$X$ compound, the doping dependence of the critical behavior near the first critical field $H_{c1}$ was studied by neutrons \cite{wulf2013} and compared to the situation in the nominally pure compound DTN \cite{wulf2015}. In contrast to the initially proposed evidence for the theoretically expected change of criticality from the BEC-type to BG-type \cite{yu2012nature}, the situation appears inconclusive: the experimentally observed critical behavior is always affected by the distribution of the critical field values and the effects of elasticity, and is probably not representative of the theoretically expected physics. We further mention the detailed neutron study of the 6\% doped DTN$X$ compound \cite{povarov2015} in which a nondispersive (local) mode is detected above the top of the magnon band. From NMR results this mode is explicitly attributed to the doped impurities \cite{orlova2017}.

\subsection{Single impurity physics}

\subsubsection{Analytical approach}\label{sec:ana_approach}

\paragraph{Single doped $S=1$ dimer.---}

\begin{figure}[ht]
    \includegraphics[width=\columnwidth,clip,angle=0]{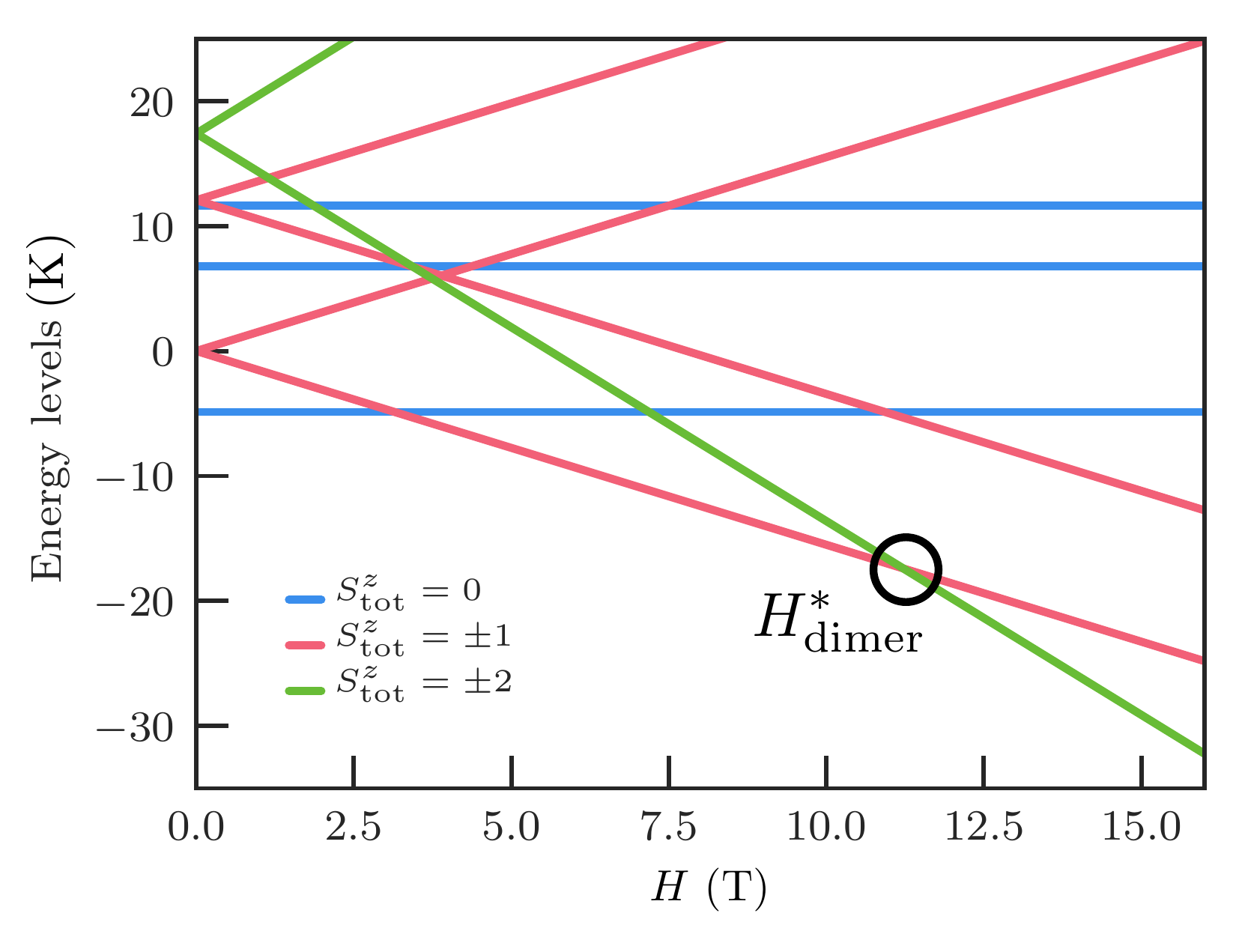}
    \caption{Energy levels of an isolated doped $S=1$ dimer  plotted against the external magnetic field. A level crossing between $S^{z}_{\rm tot}=2$ and $S^{z}_{\rm tot}=1$ states occurs at $H^{*}_{\rm dimer}\simeq 11.2$~T using realistic microscopic parameters (see text).}
    \label{fig:eigenenergy_dimer}
\end{figure}

A first step into understanding Br-doping effects is to consider a single Br impurity in a isolated $S=1$ dimer, see Fig.~\ref{fig:dtnx_sketch}(b). The resulting Hamiltonian is a $9\times 9$ block-diagonal matrix, which can be \emph{analytically} diagonalized within $S^z_\mathrm{tot}=0$, $\pm 1$ and $\pm 2$ symmetry sectors. In the following we use $J'=5.32\;\mathrm{K}$ and $D'=3.2\;\mathrm{K}$ which are the microscopic parameters determined from a direct comparison between NMR data and theory~\cite{orlova2017}. This comparison will be discussed below in Section~\ref{sec:nmrtheory}.
The eigenenergy levels  are shown in Fig.~\ref{fig:eigenenergy_dimer} against  the external magnetic field $H$. The crossing between the two lowest $S^z_\mathrm{tot}=2$ and $S^z_\mathrm{tot}=1$ levels occurs at
\bea
    H^*_\mathrm{dimer} &=& \left[J' + \frac{D'+D}{2} + \frac{1}{2}\sqrt{(D-D')^2+(2J')^2}\right]/g\mu_B\nonumber\\
    &\simeq&11.2~\mathrm{T}.
    \label{eq:DTNX_hstar_dimer}
\eea
At high magnetic field, one can restrict the problem in the vicinity of $H^*_\mathrm{dimer}$ to the two lowest-lying levels. One of them is the GS in the $S^z_\mathrm{tot}=1$ sector with eigenvector
\be
    |\Phi_1\rangle=\sqrt{\ell}|\hskip-0.1cm \uparrow \rightarrow\rangle + \rm{e}^{i\theta}\sqrt{1-\ell}|\hskip-0.1cm \rightarrow\uparrow\rangle,
\ee
where
\be
1/\ell={1+\left[\frac{D-D'}{2J'}-\sqrt{1+\left(\frac{D-D'}{2J'}\right)^2}\right]^2},
    \label{eq:epsilon_dimer}
\ee
and $\theta$ is a phase factor. The other  is the GS in the $S^z_\mathrm{tot}=2$ sector, trivially given by $|\Phi_2\rangle=$ \mbox{$|\hskip-0.1cm \uparrow\uparrow\rangle$}. The imbalance between local anisotropies, $D'\neq D$, leads to a spin imbalance between the left and right sites of the perturbed dimer. Their respective local magnetization in the $|\Phi_1\rangle$ state is simply equal to
\be
    m_z^\mathrm{left}=\ell\quad\mathrm{and}\quad m_z^\mathrm{right}=1-\ell.
\ee

Although it provides some insight to the local magnetization imbalance, this single dimer model is clearly oversimplified, as the clean environment is totally neglected. In particular, it yields a crossover field $H^{*}_{\rm dimer}\simeq 11.2$ T  {\it{below}} $H_{c2}^{\rm clean}=12.3$ T. One can easily refine this picture by adding the mean-field (MF) contribution of the surrounding spins of the clean background, assumed to be fully polarized, which leads to
\bea
H^*_\mathrm{MF}&=& H^*_\mathrm{dimer} + (J + 4J_\perp)/g\mu_B\\
&\simeq& 13\;\mathrm{T}>H_{c2}^\mathrm{clean}.\nonumber
\eea
This is self-consistent with our assumption and confirms that the clean background polarizes before the impurities in DTN$X$.

\paragraph{One impurity on a single chain.---}

\begin{figure}[ht]
    \includegraphics[width=0.8\columnwidth,clip,angle=0]{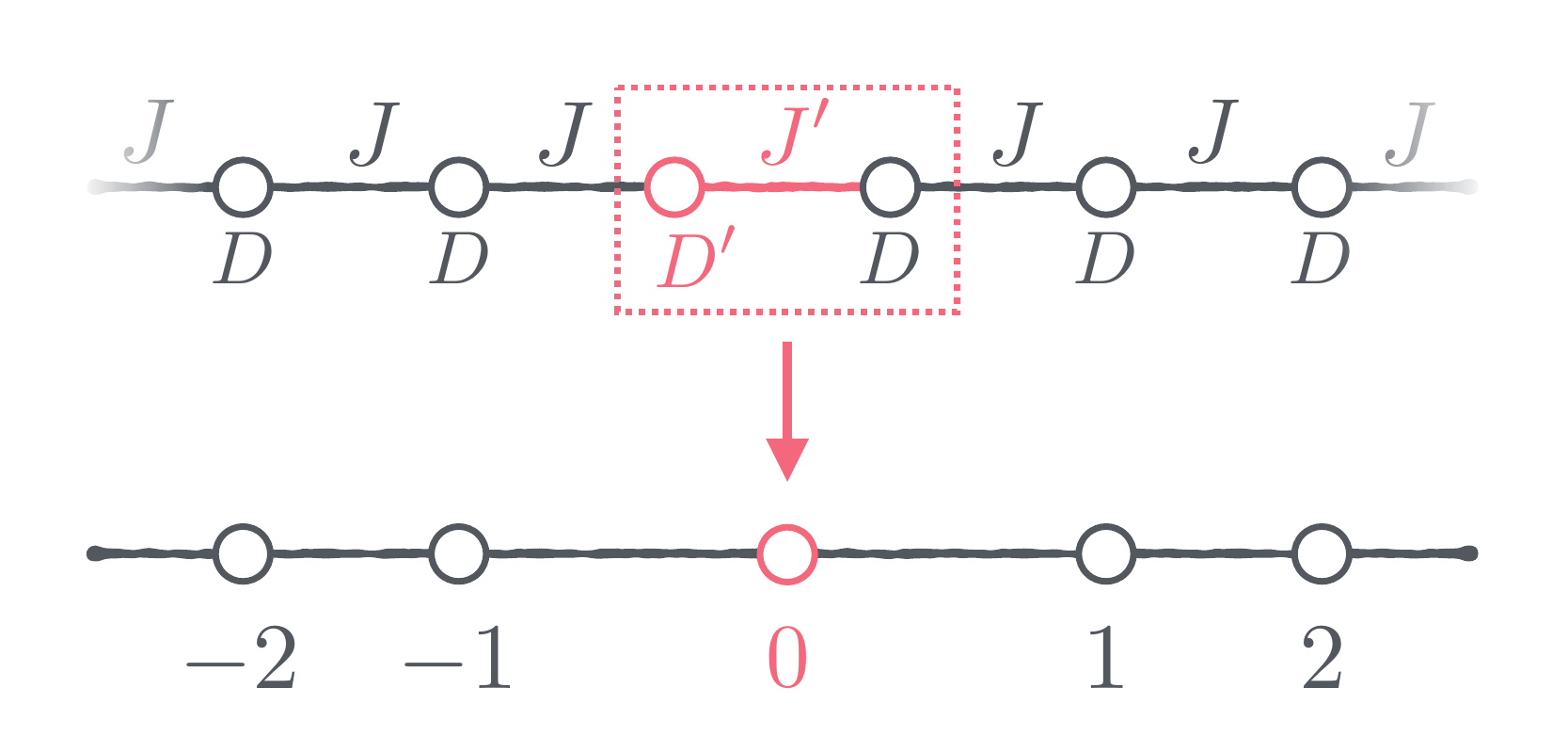}
    \caption{Effective $1$D model for the dynamics of a single impurity dimer described at high magnetic field as a two-level system $\{|\Phi_1\rangle,|\Phi_2\rangle\}$.}
    \label{fig:dtnx_effective}
\end{figure}

Going beyond the above MF scenario, we now deal with the dynamics of a single spin flipped state $|\ldots \uparrow\uparrow \rightarrow\uparrow\uparrow\ldots\rangle$ in a fully polarized background, in the presence of a central perturbed dimer. We first start this analysis on a single chain of $N$ sites and work in the $S^z_\mathrm{tot}=N-1$ symmetry sector. Using the two-level system representation $\{|\Phi_1\rangle,|\Phi_2\rangle\}$, the central dimer is replaced by a single site (at position $0$) as pictured in Fig.~\ref{fig:dtnx_effective}, which can accomodate one of the two states.

Our new basis is made of the following states labeled by the position $j$ of the flipped spin,
\bea
    |0\rangle&\equiv& |\ldots\uparrow\uparrow\uparrow\rangle |\Phi_1\rangle|\uparrow\uparrow\uparrow\ldots\rangle\nonumber\\
    |1\rangle&\equiv& |\ldots\uparrow\uparrow\uparrow\rangle |\Phi_2\rangle|\rightarrow\uparrow\uparrow\ldots\rangle\nonumber\\
    |-2\rangle&\equiv& |\ldots\uparrow\rightarrow\uparrow\rangle |\Phi_2\rangle|\uparrow\uparrow\uparrow\ldots\rangle\nonumber\\
    &&\mathrm{etc.}
\eea
In order to get a symmetric tight-binding structure for the low-energy dynamics, one has to define for $j>0$ a new set of states,
\be
    |{\overline j}\rangle=\sqrt{\ell}|j\rangle + \rm{e}^{i\theta}\sqrt{1-\ell}|-j\rangle.
    \label{eq:barj}
\ee
For an initial $S=1$ chain of $N$ sites and open boundary conditions, the dynamics in the new basis is governed by the following effective tight-binding model with ${\tilde{N}}=N/2-1$,
\bea
    \mathcal{H}_{\mathrm{tight-binding}}&=&J\sum_{j=0}^{{\tilde{N}}-1}\Bigl(|\overline{j}\rangle\langle\overline{j+1}|+|\overline {j+1}\rangle\langle\overline{j}|\Bigr)\nonumber\\
    &-&\Delta|0\rangle\langle 0| +  C\sum_{j=0}^{\tilde{N}}|\overline{j}\rangle\langle\overline{j}|,
    \label{eq:DTNX_tight_binding}
\eea
where the constant $C$ and the impurity energy shift $\Delta$ located at the ($\overline{j}=0$) boundary are respectively
\bea
    C &=&2N(D+J-H)-2J+D'+J'-H\\
    \Delta &=& J'-J+\frac{D'-D+\sqrt{(D'-D)^2+(2J')^2}}{2}\simeq 6.3\;\mathrm{K}\nonumber.
    \label{eq:imp_boundary}
\eea

Note that this description, based on the localization of the spin flip excitation on the perturbed dimer is only valid for $J'>J$.  The tight-binding Hamiltonian given by Eq.~\eqref{eq:DTNX_tight_binding}, having a localized boundary (impurity) potential $\Delta$, admits a localized GS  $|\Psi_0\rangle=\sum^{\tilde{N}}_{j=0}c_j|\overline{j}\rangle$, where $c_j\propto \exp(-j/\lambda)$ for $\Delta> J$. Inserting this form into Eq.~\eqref{eq:DTNX_tight_binding} gives

\be
    |\Psi_0\rangle=\sum^{\tilde{N}}_{j=0}c_0(-1)^j\exp\left[-j\ln\left(\frac{\Delta}{J}\right)\right]|\overline{j}\rangle.
\ee
In the limit ${\tilde{N}}\gg\lambda =1/\ln(\Delta/J)$, the occupation of the central (impurity) site is $|c_0|^2=1-\exp(-1/\xi_\parallel)$, where the localization length governing the decay of the spin density is given by
\be
    \xi_\parallel=\frac{1}{2\ln\left(\Delta/J\right)}=0.47
    \label{eq:xi_1D}.
\ee
The energy of this localized bound-state can also be obtained analytically, and the energy difference with the fully polarized state leads to the crossover field value
\bea
    H^*_{\rm 1D}&=&\left\{D+2J\left[1+\cosh\left(\frac{1}{2\xi_\parallel}\right)\right]\right\}/g\mu_B\\
    &=& H^*_\mathrm{dimer} + J/g\mu_B + J^2/(g\mu_B\Delta)\\
    &\simeq& 13.1~\mathrm{T}.\nonumber
    \label{eq:Hstar1d}
\eea
As compared to the isolated dimer picture discussed above, the first correction term corresponds to the MF contribution of the fully polarized 1D environment, $J/g\mu_B=$~1.4~T. The delocalization of the flipped spin over its neighboring sites does not extend over large scales, but it is nevertheless gains some kinetic energy, pushing the crossover field further up by $J^2/(g\mu_B\Delta)=$~0.5~T.

\paragraph{One impurity in the $3$D lattice.---}

The previous single impurity analysis can be extended to a $3$D lattice with a similar Hamiltonian to Eq.~\eqref{eq:DTNX_tight_binding}. The exponential ansatz solution now includes two different localization lengths along and perpendicular to the chain direction $\xi_{\parallel,\perp}$, with $\xi_{\parallel}$ given by Eq.~\eqref{eq:xi_1D} and
\be
    \xi_\perp=\frac{1}{2\operatorname{arcsinh}\left(\Delta/2J_\perp\right)}=0.14.
    \label{eq:xi_3D}
\ee
(The localization lengths are expressed in units of lattice spacings.)

As a result, the final crossover magnetic field is
\bea
    H^*&=&\left[H^*_\mathrm{MF}+J\mathrm{e}^{-1/2\xi_\parallel} + 4J_\perp\mathrm{e}^{-1/2\xi_\perp}\right]/g\mu_B \nonumber\\
       &\simeq& 13.6~\mathrm{T},
    \label{eq:Hstar3d}
\eea
where the very short transverse correlation length makes the last correction term negligible (0.01~T).

The magnetization profiles of the original physical (magnetic) sites at $T=0$ and for $H<H^*$ can be computed in the vicinity of the impurity. On the perturbed left and right dimer sites,
\bea
    m_z^\mathrm{left}&=&1-(1-\ell)\left[1-\mathrm{e}^{-1/\xi_\parallel}\right]\left[1-\mathrm{e}^{-1/\xi_\perp}\right]^2\,\,\,\\\label{eq:m_left}
     m_z^\mathrm{right}&=&1-\ell\left[1-\mathrm{e}^{-1/\xi_\parallel}\right]\left[1-\mathrm{e}^{-1/\xi_\perp}\right]^2,
    \label{eq:m_right}
\eea
where $\ell$ is defined in Eq.~\eqref{eq:epsilon_dimer}. A similar expression can be obtained for the magnetization of the other (clean) sites of the $3$D system.

\subsubsection{Exact Diagonalisation}\label{sec:ed_one_impurity}

\begin{figure}[t!]
    \includegraphics[width=1\columnwidth,clip,angle=0]{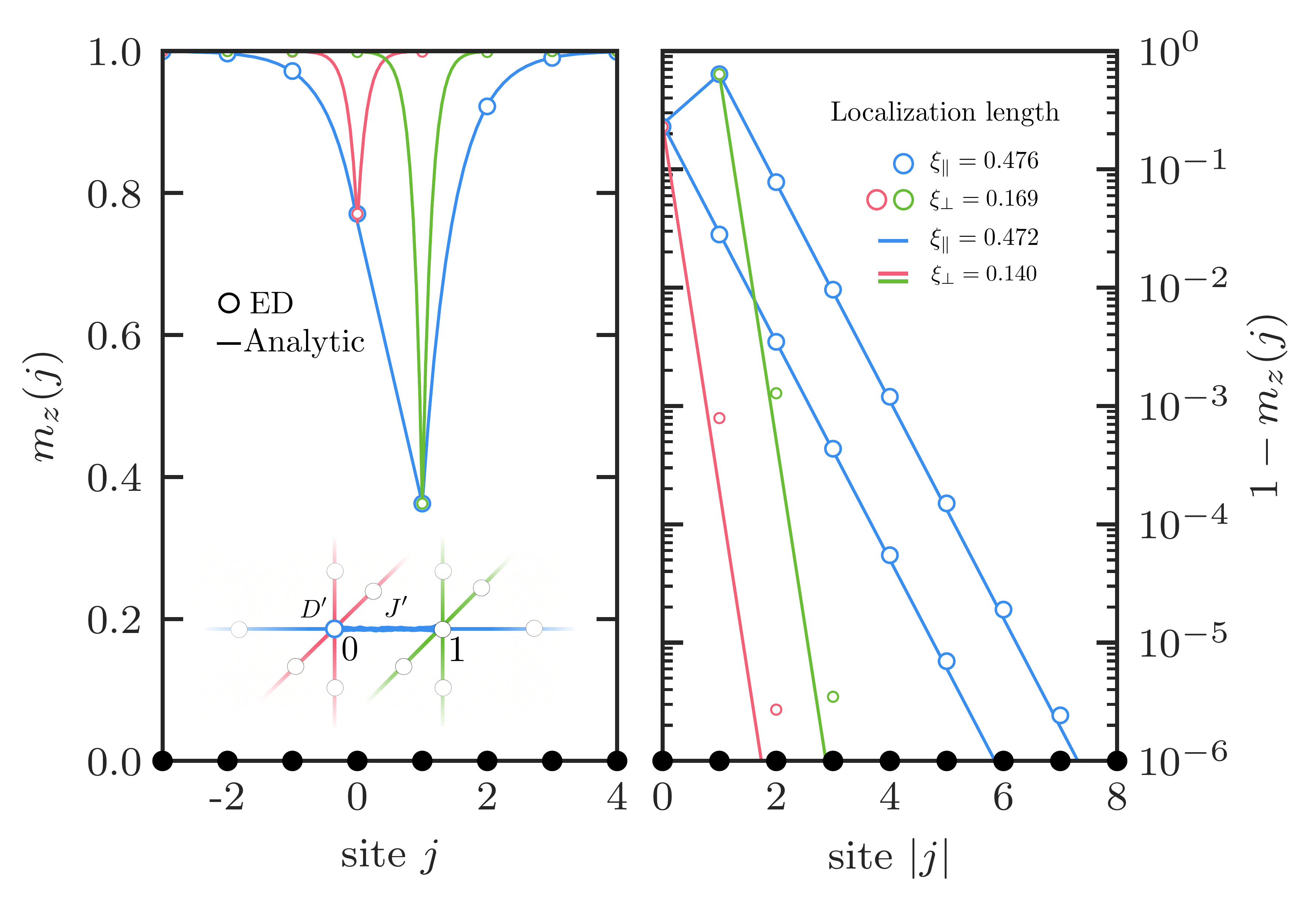}
    \caption{Local magnetization profile in the $S^z_\mathrm{tot}=N-1$ symmetry sector (single spin flip) close to a doped bond, comparing the ED results (symbols) with the analytical ones (lines). The inset defines the color code: the blue curve is along the spin chain direction and the pink/green ones are perpendicular to it. The right panel in semi-log scale shows the exponential localization of the depolarization around the impurity with very short localization lengths: $\xi_\parallel=0.476$ and $\xi_\perp=0.169$ is obtained by ED.}
    \label{fig:local_mag}
\end{figure}

Besides the analytical approach presented above for the $1$D chain and the realistic $3$D system, we also performed exact diagonalization (ED) calculations. Working in a fixed $S^z_\mathrm{tot}=N-1$ symmetry sector allows us to diagonalise large systems without much effort, the Hamiltonian matrix being of $N$$\times$$N$ size. We verified that the exponentially localized state ansatz is valid in the limit $J_\perp \ll J$, and is thus exact in the $1$D case. In Fig.~\ref{fig:local_mag} we compare the analytical results for the local magnetization with the ones computed by ED on a system of size $N=40\times 20 \times 20$ spins with one dimer located in the middle. The two results agree very well, even though in the semi-log scale one can see a small difference between the two methods, especially in the transverse direction. We also determined the correlation lengths $\xi_\parallel=0.476$ and $\xi_\perp=0.169$ by fitting ED results to exponential decays.

\subsection{NMR {\it{vs.}} theory}
\label{sec:nmrtheory}
\begin{figure}[b]
    \includegraphics[width=1\columnwidth,clip,angle=0]{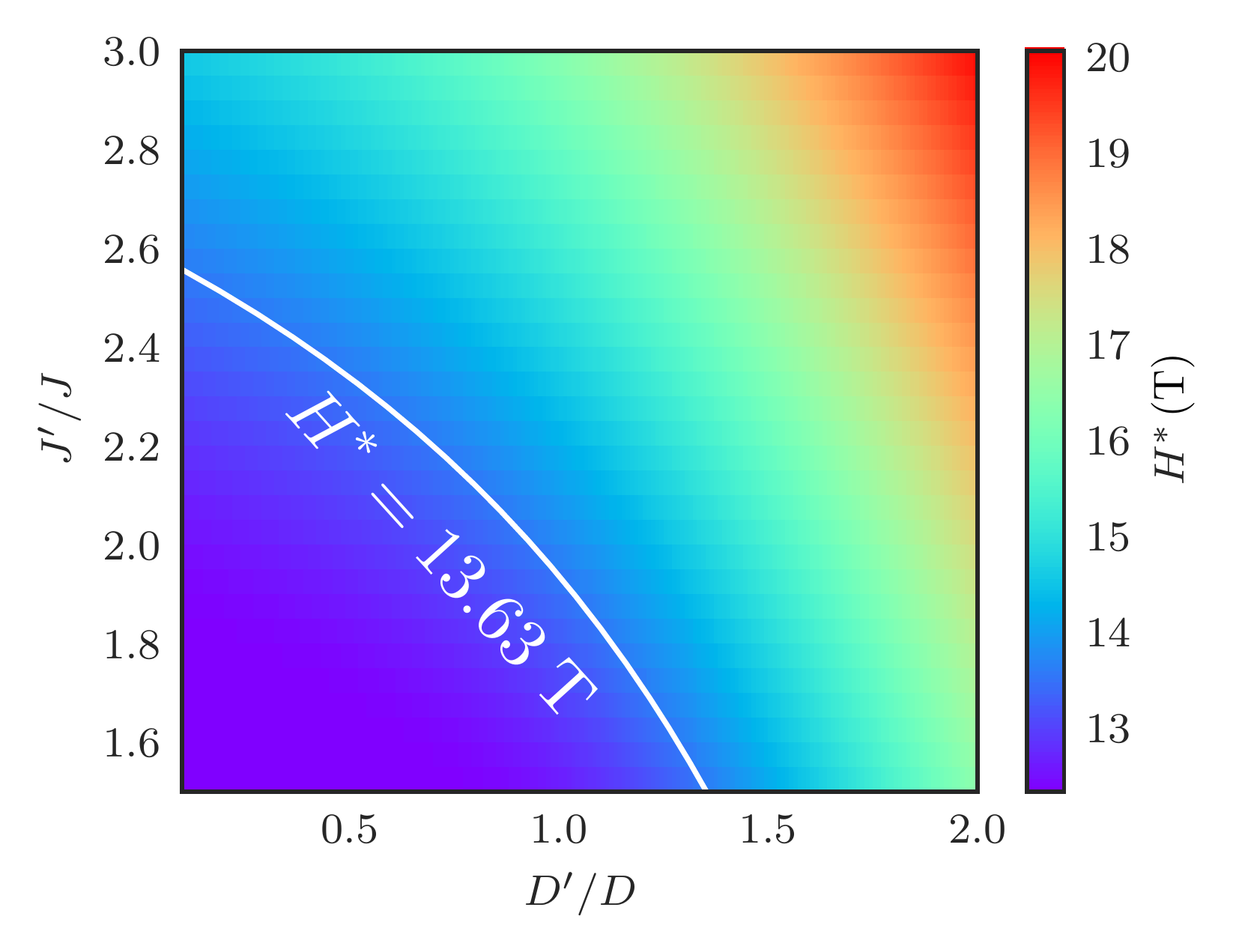}
    \caption{Color map of the crossover field $H^*$ plotted in the anisotropy--coupling plane using ED. All the pairs of parameters ($D'/D, J'/J$) along the white line correspond to the same value of the crossover field $H^*=13.63\;\mathrm{T}$.}
    \label{fig:hstar_colormap}
\end{figure}
The above described level crossing is clearly evidenced in the recent NMR results presented in Ref.~\onlinecite{orlova2017}. Before discussing these, we first recall the archetypal signature of a level crossing as observed in molecular crystals consisting of antiferromagnetic spin rings: the molecular level crossing is there observed as a sharp, tanh-shaped step in the magnetization that is concomitant with a peak of the $T_1^{-1}$ NMR relaxation rate, whose magnetic field dependence at low temperature directly reflects the corresponding linear opening of the gap between the two levels \cite{julien1999,micotti2005}. In DTN$X$ the NMR data \cite{orlova2017} also show a peak in $T_1^{-1}$ at the same field value, $H^*=13.63\;\mathrm{T}$, where a step was previously observed in the bulk magnetisation data \cite{yu2012nature}. The position ($H^*$) of this $T_1^{-1}$ peak is found to be nearly doping independent, which means that it should be associated to a single-impurity effect. Furthermore, the field dependence of the $T_1^{-1}$ peak reveals the linear gap opening above $H^*$, thereby confirming the level crossing scenario. (We cannot use the dependence observed below $H^*$ because it is affected by the critical behaviour related to the nearby QPT at $H_{c2}$.)

\begin{figure}[t!]
    \includegraphics[width=1\columnwidth,clip,angle=0]{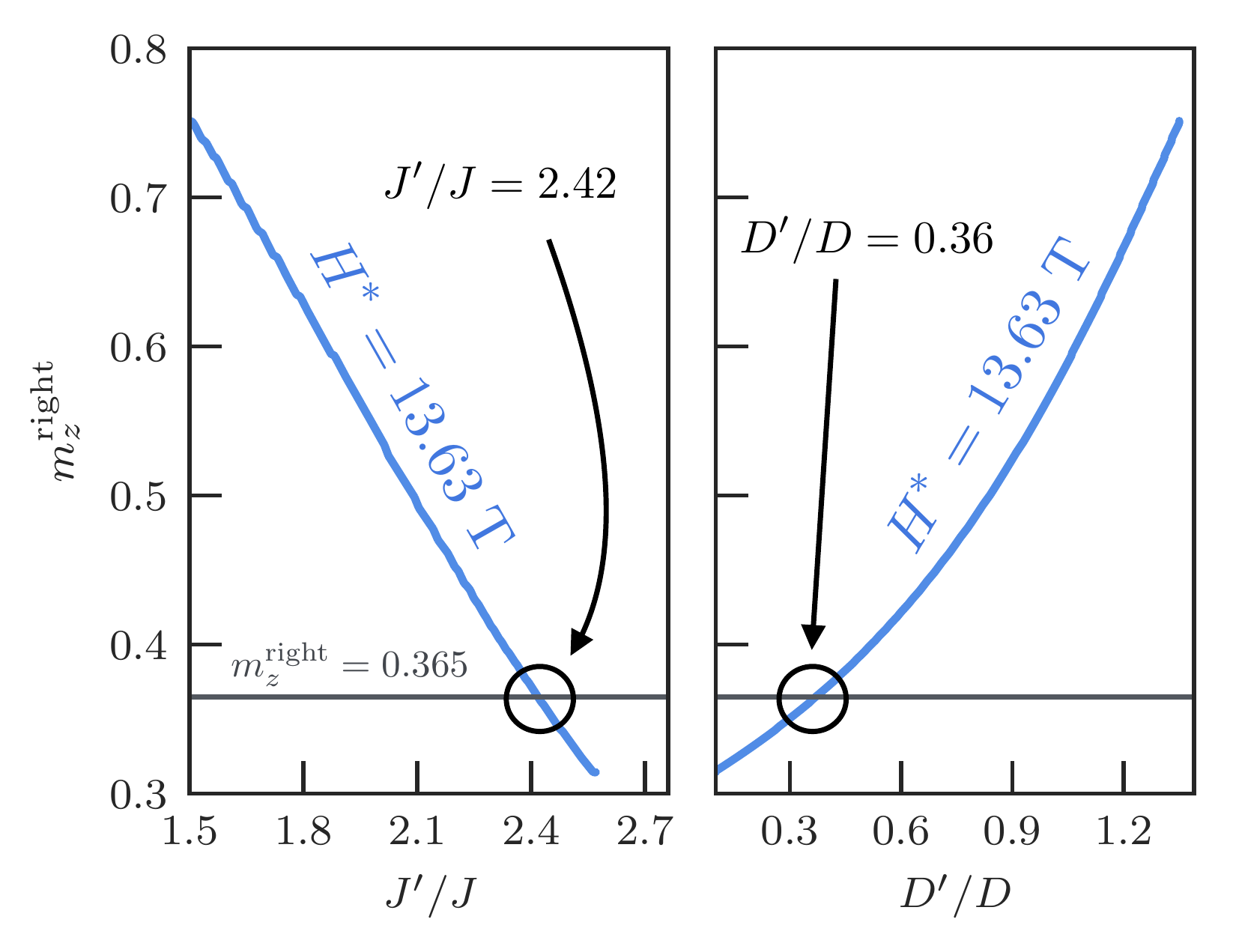}
    \caption{Zero temperature magnetization of the right-hand-side site of the doped dimer below $H^*$, calculated using ED, plotted against one of the adjustable parameters $J'$ and $D'$, while the other parameter is adjusted to maintain the crossover field at $H^*=13.63\;\mathrm{T}$ (see the white line in Fig.~\ref{fig:hstar_colormap}). The bottom grey line shows the experimental value of $m_z^\mathrm{right}=0.365$.}
    \label{fig:local_mag_hstar}
\end{figure}

The NMR spectra provided the second key-information to describe the impurity levels: the precise value of the local polarization of the spin at the right-hand-side of the dimer, as sketched in Fig.~\ref{fig:dtnx_sketch}(b) and labeled as ``site $1$'' in Fig.~\ref{fig:local_mag}. Below $H^*$ and at low temperature this site is depolarized to $m_z^\mathrm{right}=0.365$, which provides the second independent information on the impurity states. Together with the $H^*=13.63\;\mathrm{T}$ value determined from the position of the $T_1^{-1}$ peak at low temperature, using equations \eqref{eq:Hstar3d} and \eqref{eq:m_right}, or, equivalently, the ED results shown in Figs.~\ref{fig:hstar_colormap} and \ref{fig:local_mag_hstar}, we can precisely determine the two local impurity values $D'$ and $J'$,
\be
   J'=2.42J\quad\mathrm{and}\quad D'=0.36D.
   \label{eq:DTNX_parameters}
\ee
These values are in agreement with the ones proposed previously ($J'=2.35J$ and $D'=0.5D$) \cite{yu2012nature} from the global fits, where the determination was mostly relying on the $H^*$ value only (as plotted in Fig.~\ref{fig:hstar_colormap}), so that the $D'$ value was in fact not precisely known.

Finally, the NMR results \cite{orlova2017} provide also clear evidence of effects going beyond the single-impurity description: the temperature dependence of the level-crossing gap above $H^*$ reveals that the gap value is (inhomogeneously) distributed, and the local polarization $m_z^\mathrm{right}$ above $H^*$ is found to present an unexpected field dependence at low temperature. Furthermore, a weak secondary peak of $T_1^{-1}$ was found above $H^*$ at $H^{**}=15.2$~T. This brings us to the following section that treats the many-body effects.

\section{Many-body effects beyond single impurity}\label{sec:manybody_impurity}

\subsection{Effective theory from pairwise interactions}

\subsubsection{Mutual effect of two impurities}

The above given analysis of a single Br-doped bond provides us with a precise picture of DTN\textit{X} above $H_{c2}$: the clean background is fully polarized and only the sites in the direct vicinity of Br-impurities remain not yet fully polarized, whereas this depolarization is exponentially localized. The localization lengths in both longitudinal and transverse directions are way shorter than one lattice spacing unit. In realistic DTN\textit{X} samples with low doping concentration, $2x\ll 1$, isolated impurities (of ``length'' $l=1$) are the most common objects. However, there are also other objects, zones or clusters consisting of more than one isolated impurity ($l>1$). As long as $2x<31.2\%$ --- the site percolation threshold on a cubic lattice~\cite{deng2005} --- there cannot be an infinite-size Br-doped cluster in the sample. Below this limit, plethora of impurity clusters of various sizes and spatial configurations may exist, but the bigger ones are more rare. Moreover, the bigger they are, the larger the magnetic field value has to be to polarize them, giving rise to Lifshitz tails~\cite{lifshitz1964,friedberg1975} in the magnetization curve, up to $H'_{c2}=(D'+4J'+8J_\perp)/g\mu_B\simeq 16.7$~T, which is the second critical field for the hypothetical homogeneously and fully doped sample, $2x=1$. Above $H'_{c2}$ all the impurity cluster sizes, and thus the whole sample, are necessarily totally polarized, as shown in Fig.~\ref{fig:phase_diagram_3d}. In the following, we will first consider the mutual effect of \textit{two} impurities depending on their relative distance $r_{\parallel,\perp}$.

\begin{figure}[b!]
    \includegraphics[width=1\columnwidth,clip,angle=0]{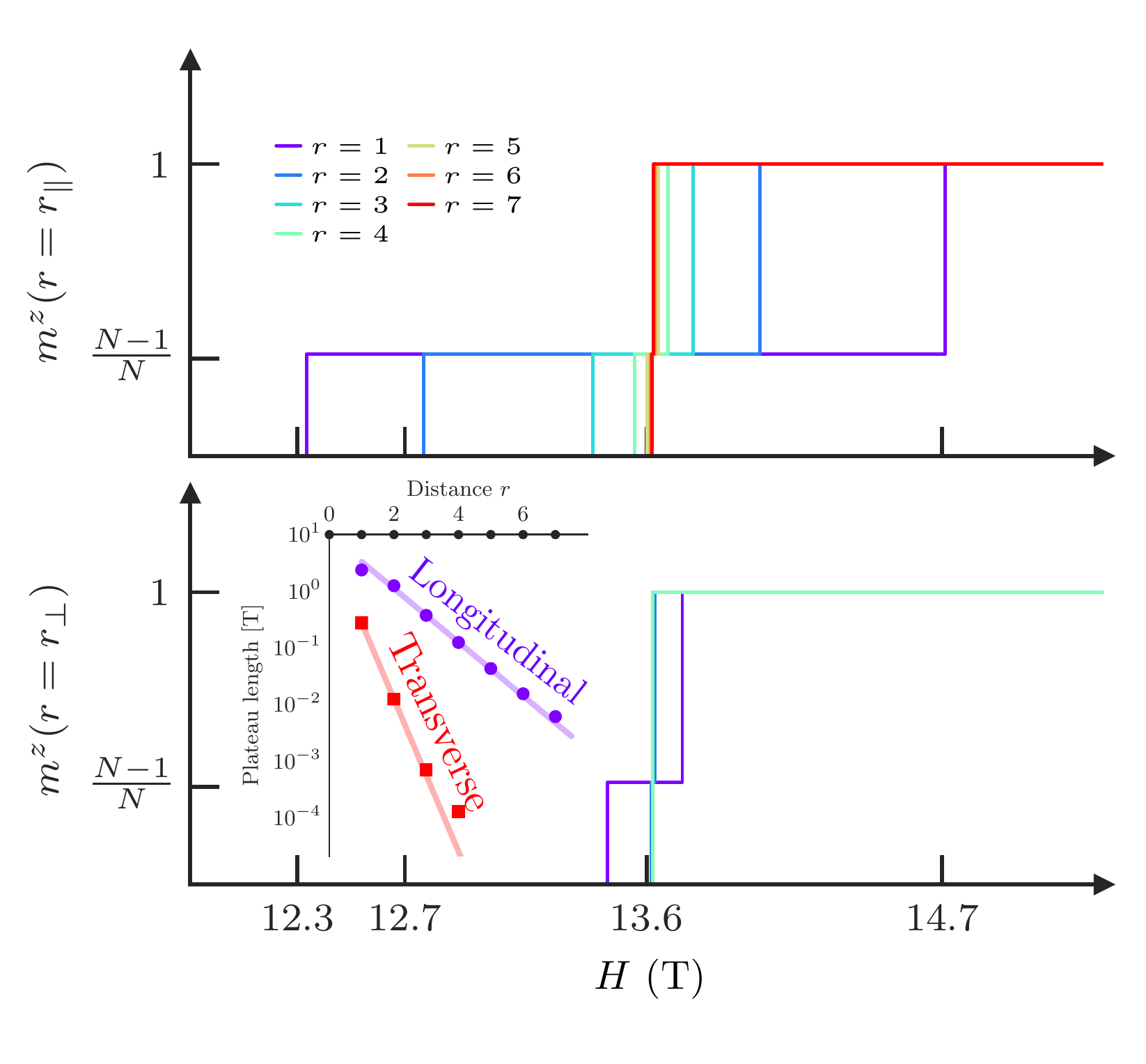}
    \caption{ED results for the magnetization steps in a $N=16\times 8\times 8$ system with two impurities. Top: impurities in the same chain at distance $r=r_\parallel$. Bottom: impurities in different chains at distance $r=r_\perp$ ($r_\parallel=0$). In both cases, as a function of inter-impurities distance $d$, a magnetization plateau develops at $m^z=(N-1)/N$ when impurities get closer. Inset: Behavior of the plateau \textit{length} as a function of the distance $r$ (symbols). An exponential decay $\sim \exp(-r_{\parallel,\perp}/\lambda_{\parallel,\perp})$ (lines) is observed, where $\lambda_\parallel\simeq 0.92$ and $\lambda_\perp\simeq 0.32$. }
    \label{fig:mag_two_imp}
\end{figure}

To this end we performed ED computations in the high magnetization sectors $S^z_\mathrm{tot}=N, N-1, N-2$ of the $3$D system described by Eq.~\eqref{eq:DTNX} containing $N=16\times 8\times 8$ spins and two impurities located at varying distances $r_{\parallel,\perp}$. The magnetization process of the two impurities is shown in Fig.~\ref{fig:mag_two_imp} for increasing distances $r_{\parallel,\perp}$. For short relative separation between two impurities, a  magnetization plateau at $S^z_\mathrm{tot}=N-1$ is clearly visible. However, its width gets rapidly reduced when the two dopants are moved apart. When $r_{\parallel,\perp}$ is large enough, the plateau width shrinks to zero, and one recovers the already discussed single impurity limit: a single level crossing at $H^*=13.63\;\mathrm{T}$ between the GS energy of the $S^z_\mathrm{tot}=N$ and $S^z_\mathrm{tot}=N-2$ symmetry sectors.

The presence of such plateaus at short distances is a signature of the mutual effect of the two impurities. In the inset of Fig.~\ref{fig:mag_two_imp} an exponential decay for the size of these plateaus is reported as a function of the relative distance between impurities for both parallel and perpendicular directions. The length scales controlling such decay reflect the localization lengths $\lambda_\parallel\simeq 0.92\sim 2\xi_\parallel$ and $\lambda_\perp\simeq 0.32\sim 2\xi_\perp$. We further study this exponential decay of the effective coupling between impurity states in the next subsection.

\subsubsection{Effective bosonic description}\label{sec:effective_model}

Having realized that close-by impurities do not behave as isolated, it becomes clear that many-body physics should play a role in DTN$X$, and that one has to consider the pairwise effects. We therefore propose an effective hard-core bosons (HCB) model description for DTN$X$ at high magnetic field ($H>H_{c2}$), based on ED and which reveals an effective AF pairwise interaction between the impurities around $H^*$. Again, what is called an impurity corresponds to a Br-doped bond as pictured in Fig.~\ref{fig:dtnx_sketch}(b), which exponentially localizes the depolarization.

The picture of the effective model is as follows: the fully polarized state is the vacuum and decreasing the field will lead to more and more depolarized impurities, which we take for the effective particles. The initial model of DTN$X$ is mapped to an HCB model where the number of particles is controlled by a chemical potential (magnetic field). The size of the local Hilbert space (labeled by $|1\rangle$ and $|0\rangle$) is therefore reduced to the presence or not of a particle, or, in the initial language, to a depolarized or polarized impurity. The most generic HCB hamiltonian limited to a two-body interaction is
\bea
    \mathcal{H}_\mathrm{tV}&=&\sum_{i,j}\left[t_{ij}\left(b^\dag_ib_j + \mathrm{h.c.}\right)+ V_{ij}n_in_j\right]\nonumber\\
    &-&\sum_i\mu_in_i+ C,
    \label{eq:effective_ham}
\eea
where $t_{ij}$ is the hopping strength, $V_{ij}$ is the interaction potential and $\mu_i$ is the chemical potential. $C$ is a constant shift of the whole energy spectrum. The operators $b_i^\dag$ and $b_i$ are respectively the creation and annihilation operators of HCB ($\langle n_i\rangle=\langle b^\dag_ib_i\rangle\leq 1$) on site $i$. They obey bosonic commutation relations $[b_i,b_j^\dag]=0$ on different sites $i\ne j$ and fermionic ones on the same site $\{b_i,b_i^\dag\}=1$. The summation is over all possible sites $i\neq j$ containing an impurity in the initial model. The idea is to determine the Hamiltonian parameters that will reproduce the most faithfully the way impurities (de)polarize, taking into account the many-body effects.

To obtain the effective model parameters, we project the wave-functions of the low-energy spectrum of the real DTN$X$ model Eq.~\eqref{eq:DTNX} onto the effective model. Since we have a pairwise interaction between the particles, we perform these calculations with two impurities at positions $i$ and $j$ in the initial spin $S=1$ model varying the distance between them in the longitudinal ($r_\parallel$) and the transverse ($r_\perp$) directions. We use ED on the initial model in $S^z_\mathrm{tot}=N$, $N-1$ and $N-2$ symmetry sectors and make the following correspondence between the states of initial model and the effective one. First, the vacuum is associated to the fully polarized state $|\varphi_{N}\rangle$ of energy $E_N$ and defines the energy shift $C$,
\be
    E_N|\varphi_{N}\rangle~\longrightarrow~C|0_i0_j\rangle.
\ee
Then, we associate the state with two particles in the effective model with the $S^z_\mathrm{tot}=N-2$ symmetry sector GS $|\varphi_{N-2}\rangle$ of energy $E_{N-2}$,
\be
    E_{N-2}|\varphi_{N-2}\rangle~\longrightarrow~(V_{ij} - \mu_i - \mu_j + C)|1_i1_j\rangle.
\ee
The correspondence in the $S^z_\mathrm{tot}=N-1$ symmetry sector is a bit more sophisticated as we have two possible different states $|0_i1_j\rangle$ and $|1_i0_j\rangle$ in the effective model. Considering the dimer states $|\Phi_1\rangle$ and $|\Phi_2\rangle$ defined in Sec.~\ref{sec:ana_approach}, we build the following two states in the initial spin language,
\bea
    |\phi_1\rangle &=& |\uparrow\uparrow\uparrow\cdots\rangle|\Phi_2\rangle|\uparrow\uparrow\uparrow\cdots\rangle|\Phi_1\rangle|\uparrow\uparrow\uparrow\cdots\rangle,\nonumber\\
    |\phi_2\rangle &=& |\uparrow\uparrow\uparrow\cdots\rangle|\Phi_1\rangle|\uparrow\uparrow\uparrow\cdots\rangle|\Phi_2\rangle|\uparrow\uparrow\uparrow\cdots\rangle,
    \label{eq:sztot_minus_one}
\eea
where $|\Phi_1\rangle$, $|\Phi_2\rangle$ are at the positions $i$ and $j$ of the two impurities. We assume that linear combinations of $|\phi_1\rangle$ and $|\phi_2\rangle$ will be good approximations of the exact states $|\varphi_{N-1}\rangle$ (GS of energy $E_{N-1}$) and $|\varphi'_{N-1}\rangle$ (first excited state of energy $E'_{N-1}$) of the $S^z_\mathrm{tot}=N-1$ symmetry sector. These exact states are projected onto the trial ones,
\bea
    |\psi_1\rangle &=& |\phi_1\rangle\langle\phi_1|\varphi_{N-1}\rangle+|\phi_2\rangle\langle\phi_2|\varphi_{N-1}\rangle,\nonumber\\
    |\psi_2\rangle &=& |\phi_1\rangle\langle\phi_1|\varphi'_{N-1}\rangle+|\phi_2\rangle\langle\phi_2|\varphi'_{N-1}\rangle,
    \label{eq:sztot_minus_one2}
\eea
which are orthogonalized using standard Gram-Schmidt procedure and normalized to form a new eigenbasis with respective energies $E_{N-1}$ and $E'_{N-1}$. We then make the correspondence between the effective Hamiltonian matrix in the basis $\{|1_i0_j\rangle,|0_i1_j\rangle\}$ and the initial model,
\bea
    &&E_{N-1}|\psi_1\rangle\langle\psi_1|\nonumber\\
    &+&E'_{N-1}|\psi_2\rangle\langle\psi_2|~\longrightarrow~
    \begin{pmatrix}
        C - \mu_i & t_{ij} \\
        t_{ij} & C - \mu_j
    \end{pmatrix}.
    \label{eq:sztot_minus_one3}
\eea
When the two impurities are spatially well separated with no overlap of the exponentially localized depolarization clouds, the GS is twice degenerated as expected, with $E_{N-1}=E'_{N-1}$ when $r_{\parallel,\perp}\gg 1$.

\begin{figure}[ht]
    \includegraphics[width=1\columnwidth,clip,angle=0]{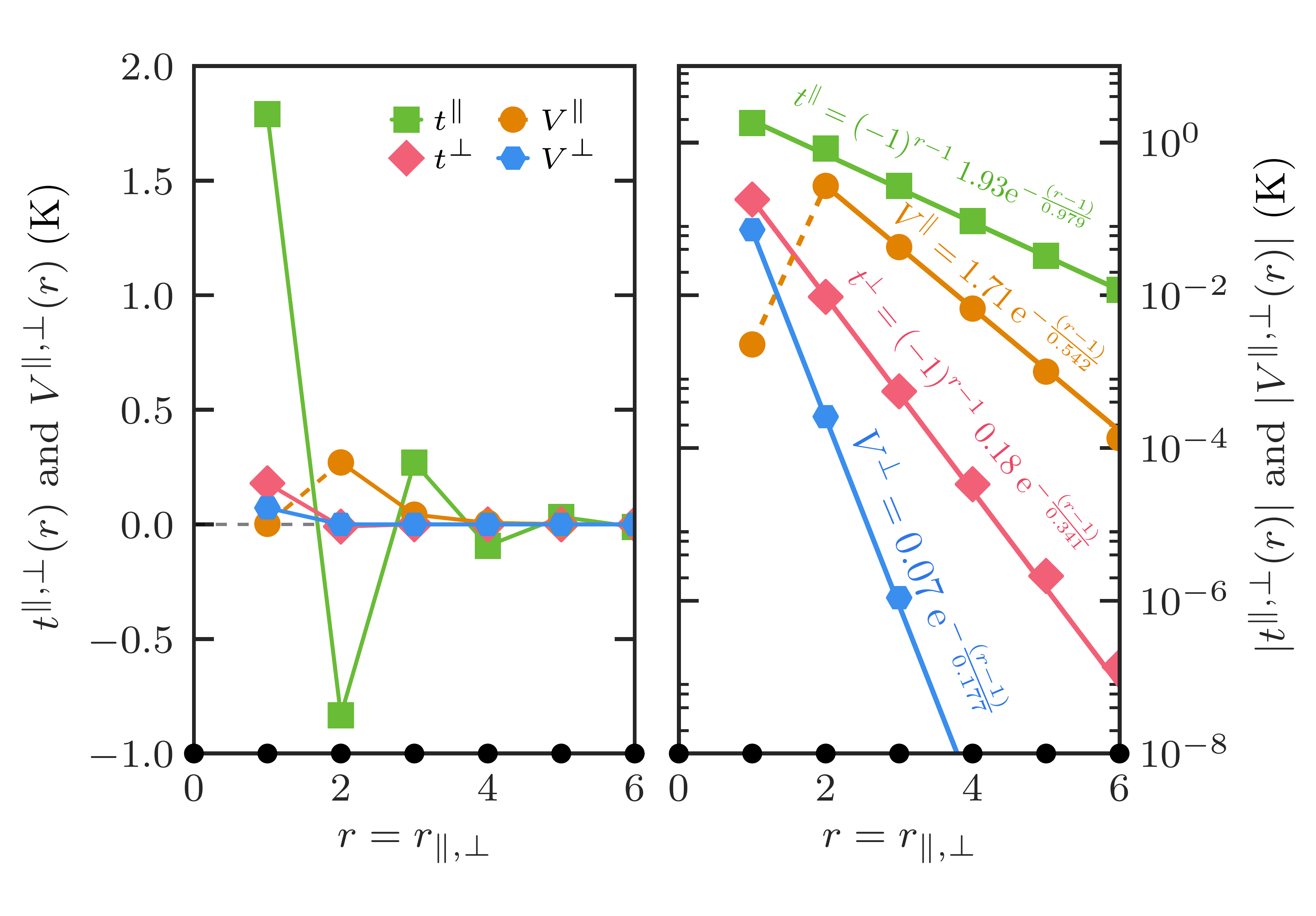}
    \caption{Effective coupling parameters defined by Eq.~\eqref{eq:effective_ham} and determined using ED, plotted as a function of the distance $r_{\parallel,\perp}$ between the impurities. The results are given in both linear (left panel) and logarithmic scale (right panel). The hopping term $t$ displays the AF character of the underlying microscopic model in both the longitudinal (green squares) and transverse (pink diamonds) directions. The interaction potential, although non-staggered, decays more rapidly with distance than the hopping terms (yellow circles in the longitudinal direction and blue hexagons in the transverse one).
    }
    \label{fig:effective_model}
\end{figure}

The above procedure fully determines the parameters of the effective Hamiltonian. They are computed varying the distance between the two impurities along the main chain ($r_\parallel$) and the perpendicular direction ($r_\perp$) and plotted in Fig.~\ref{fig:effective_model}.
ED calculation is performed on a system of $N=16\times 8\times 8$ spins with periodic boundary conditions. The hopping term $t$ and the interaction potential $V$ are both exponentially decaying:
\be
t,V\propto \exp\left(-|\mathbf{r}|/\lambda\right)\ee
 with
 \be\lambda_{\parallel,\perp}\simeq 2\xi_{\parallel,\perp}\,\,
 {\rm{for}}\,t\,\,\,\,\,
 {\rm{and}}\,\,\,\,\,
 \lambda_{\parallel,\perp}\simeq \xi_{\parallel,\perp}\,\,
 {\rm{for}}\, V,\ee
  where $\xi$ is the localization length of the wave-function around the impurity introduced in Eqs.~\eqref{eq:xi_1D} and~\eqref{eq:xi_3D}. The hopping parameter is non-frustrated and preserves the AF character of the underlying microscopic model. The interaction potential is frustrated but decays more quickly than the hopping term, making it typically one or more orders of magnitude smaller. We thus assume this frustrated term to be irrelevant and therefore neglect it in the following. The chemical potential value is site-independent, with $\mu_i=g\mu_B(H-H^*)=\mu$ which controls the density of particles. Shifting the Hamiltonian~\eqref{eq:effective_ham} by $-C$, we finally get
\bea
    \mathcal{H}_\mathrm{eff}&=&\sum_{i,j}t_{ij}\left(b^\dag_ib_j + \mathrm{h.c.}\right)-\mu\sum_i n_i.
    \label{eq:effective_DTNX}
\eea

This effective HCB model gives a quite simple two-level system description for the localized states living in the vicinity of Br-impurities. One should emphasize that\\
(i) This effective Hamiltonian [Eq.~\eqref{eq:effective_DTNX}] is defined on a sparse 3D network of $2x\times N$ active sites.\\
(ii) These sites are coupled through {\it{non-frustrated}} hopping terms which decay exponentially with their relative separation, yielding a random hopping problem due to the random location of the impurities in the original 3D cubic lattice.\\
(iii) The HCB density is controlled by a non-random chemical potential $\mu\propto H-H^* $.

The interplay between disorder and interactions for bosonic systems has mostly been investigated for diagonal disorder, i.e. random potentials. Here the disorder is of off-diagonal nature, i.e. with random hopping $t_{ij}$, a problem which has been less studied. At the single-particle level, it is known that randomness of the hopping terms modifies the Anderson localization at the center of the band where a delocalized state exists~\cite{eggarter1978,soukoulis1982}. Moreover, for the so-called Lifshitz model~\cite{lifshitz1965}, describing 3D diluted semiconductors with isotropic hopping terms $\sim \exp(-r/\xi)$, it was shown that extended states exist if the impurity density is above the critical one, $\rho> \rho_c$, where $\rho_c\simeq (3\xi)^{-3}$~\cite{ching1982}.

In the presence of interactions,  a few existing studies of random exchange quantum antiferromagnets have shown that long-range order remains in the presence of disorder~\cite{laflorencie2004,laflorencie2006}, a phenomenon corroborated by order-from-disorder mechanisms observed in quantum spin gapped materials doped with impurities~\cite{wessel2001,bobroff2009}.

In view of these results, one can reasonably expect a similar effect for the effective Hamiltonian [Eq.~\eqref{eq:effective_DTNX}], at least in the vicinity of half-filling, when $H\sim H^*$. More precisely, the effective bosonic degrees of freedom, hopping on a diluted $3$D lattice, should display low temperature long-range order, i.e. BEC, meaning transverse magnetic order for the original DTN$X$ material. This general expectation has been unambiguously confirmed by realistic simulations~\cite{dupont2017}. However, this description is limited to strong dilution $2x\ll 1$, where many-body physics is faithfully captured by the above pairwise coupling approach. Below, we go beyond and address the question of multi-impurity effects which may modify this simple picture.

\subsection{Multi-impurity effects}
\label{sec:mie}
\subsubsection{Magnetization curve}

\begin{figure}[t!]
    \includegraphics[width=1\columnwidth,clip,angle=0]{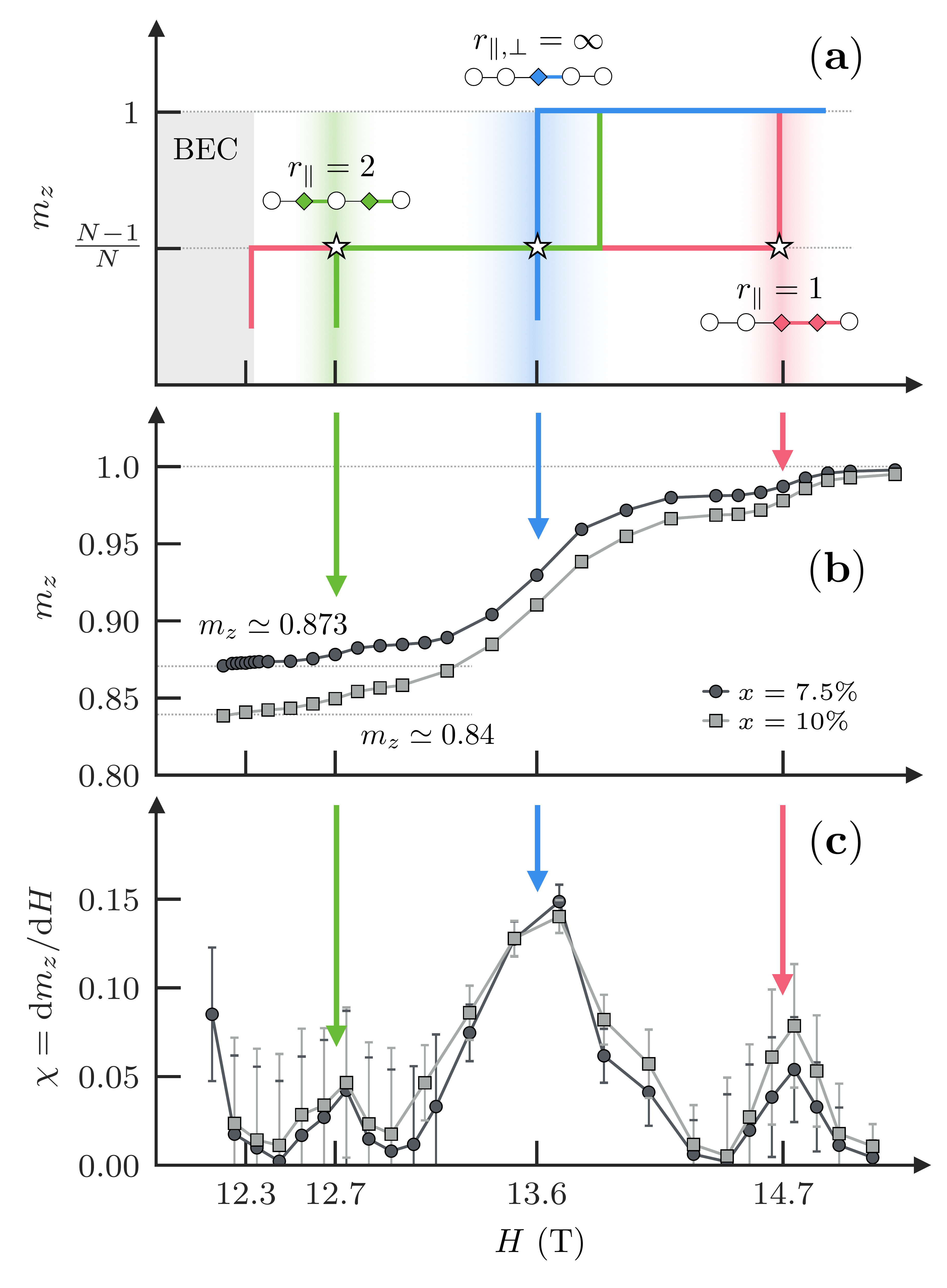}
    \caption{The top (a) panel is a simplified version of Fig.~\ref{fig:mag_two_imp}, focussing on the clearly separated level crossings for two impurities at distance $r=1,2$ and $\infty$. The central panel (b) displays the zero temperature magnetization curve at doping concentration $x=7.5\%$ (circle) and $x=10\%$ (square) from numerical simulations of the Hamiltonian~\eqref{eq:DTNX} using QMC and the $\beta$-doubling scheme as explained in Sec.~\ref{sec:qmc_bdoubling}. The results are from samples containing $N=40\times 8\times 8$ spins and each point is averaged over $200$ disorder configurations. The bottom panel (c) shows the first numerical derivative of the magnetization curve (corresponding to the magnetic susceptibility $\chi$). The highlighted levels crossing in the first panel are clearly identified in realistic numerical simulations, as denoted by vertical arrows.}
    \label{fig:magnetization}
\end{figure}

The zero-temperature high magnetic field magnetization profile of DTN$X$, obtained using QMC simulations, is shown in Fig.~\ref{fig:magnetization}(b) where step-like features are clearly visible. To understand this dependence, we focus on objects made of two impurities ($l=2$) close to each others at distances $r_\parallel=1,2$ along the chain direction, and which happen with respective probabilities $\propto(2x)^2$ and $\propto(1-2x)(2x)^2$. Longer distances rapidly approach the isolated impurity case ($r_\parallel\gg 1$), as shown in Fig.~\ref{fig:mag_two_imp}. Furthermore, the case of impurities next to each others in the transverse directions can be considered as practically isolated impurities due to the extremely short localization length $\xi_\perp$. Finally, considering objects consisting of more than two impurities ($l\geq3$) is equivalent to deal with rare events due to the very small probability of existence, $\propto x^l$.

First of all, as pictured in Figure~\ref{fig:magnetization}~(a), the $r_\parallel=1$ case has a first level crossing around $H=12.3\;\mathrm{T}$, close to $H_{c2}$, the critical field which ends the BEC phase. This provides a very simple explanation for the measured magnetization both experimentally by Yu {\it{et al.}}~\cite{yu2012nature} and in numerical simulations presented in Fig.~\ref{fig:magnetization}~(b) where $m_z$ is found to be larger than $\simeq (1-2x)$, the value expected  if only clean sites were fully polarized. Instead, at $H_{c2}$ impurities in this particular $r_\parallel=1$ configuration are (half-)polarized which leads to a total magnetization $m_z\simeq(1-2x)+(2x)^2$, in excellent agreement with both experimental results~\cite{yu2012nature} and our numerical simulation shown in Fig.~\ref{fig:magnetization}~(b). Also, instead of sharp, square steps sketched in Fig~\ref{fig:magnetization}(a), in a realistic sample one expects smooth, rounded steps at the edge of the plateaus, due to the bandwidth of the levels crossing, resulting from the effective interaction between impurities. Moreover, something like true plateaus cannot exist, because there is a multitude of levels crossing corresponding to all the different impurity cluster configurations. We are therefore left with a compressible (non-zero magnetic susceptibility) phase up to $H'_{c2}$. Although this phase is compressible, level crossings at $H=12.7\;\mathrm{T}$, $13.6\;\mathrm{T}$ and $14.7\;\mathrm{T}$ stand out because (i) they are well isolated from the others and (ii) concern a relatively large number of objects, which explains the strong step-like feature visible at these specific magnetic fields in the magnetization curve [Fig.~\ref{fig:magnetization}(b)] or in its first derivative, the magnetic susceptibility curve [Fig.~\ref{fig:magnetization}(c)]. In other words, at each of these levels crossing, there is a qualitative change in the sample as a macroscopic number of the studied ``impurity objects'' are getting polarized at the same time. As already mentioned for the $r_\parallel=\infty$ case, the simultaneous closing of the local gap $\Delta(r_\parallel=1,2,\infty)$ of these objects, together with an effective pairwise AF interaction between them,
opens the door to a global phase coherence of these new objects, in sharp contrast with the BG regime predicted in Ref.~\cite{yu2012nature}. This scenario was indeed verified in Ref.~\onlinecite{dupont2017} around $H^*=13.6\;\mathrm{T}$ where a BEC$^*$ of the single impurities was numerically observed. Based on similar mechanisms, we claim that there should also exist LRO at the two other levels crossing, $H=12.7\;\mathrm{T}$ and $H=14.7\;\mathrm{T}$, with possible intermediate BG regions at low doping concentrations as sketched in Fig.~\ref{fig:phase_diagram_3d}. Before presenting numerical evidence for such a scenario below in Section~\ref{sec:DTNX_physics}, we address now the experimental facts concerning multi-impurity physics in DTN$X$.

\subsubsection{Experimental evidences}

The first experimental evidence for the level crossing at 12.7~T can be found in the ac-susceptibility data presented in Fig.~2(b) of Ref.~\cite{yu2012nature}: there is a barely visible peak in the magnetic field dependence, present only in the lowest temperature, 1~mK data. Recent NMR data \cite{orlova2017} provide a clear direct experimental evidence for the level crossing at 14.7~T: the low temperature (113 mK) data for the $T_1^{-1}$ relaxation rate of protons present a clear peak at slightly higher field value $H^{**}=15.2$~T. The small difference between the predicted and the observed $H^{**}$ value can be accounted for by improving the above given simplest model that describes DTN$X$ by the first trivial correction: the impurity modifying the $J$ coupling into $J'$ value corresponds only to the most probable configuration where the doped bond is between one affected spin, having the anisotropy $D'$, and one unaffected spin, having the ``normal'' anisotropy $D$. When two \emph{neighboring} bonds are  doped, the bond between the two affected spins, both having $D'$ anisotropy, is in fact expected to have somewhat different exchange coupling $J''$. Indeed, a slight modification, $J''=1.12J'$, is enough to match the theoretically predicted value with the experimentally observed $H^{**}$. We have thus clearly explained the observed peak of $T_1^{-1}$ and quantified the first obvious correction to the model. This correction being small, for simplicity, we have neglected it in numerical simulations.

\section{Impurity-induced long-range order at finite temperature}\label{sec:DTNX_physics}

\subsection{Quantum Monte-Carlo simulations}

We use QMC through the stochastic series expansion (SSE) algorithm~\cite{syljuaasen2002,bauer2011} to simulate the DTN$X$ Hamiltonian~\eqref{eq:DTNX}. Simulations are performed on $3$D systems of $N=L\times L/R\times L/R$ sites, where $R>1$ is an anisotropic aspect ratio~\footnote{We used $(x,R)=~(16.67\%,6)$, $(12.5\%,8)$, $(10\%,10)$ for data obtained through ``standard QMC'' and $(x,R)=~(10\%,5)$, $(7.5\%,5)$ for data obtained through the QMC $\beta$-doubling scheme.}, numerically favorable~\cite{sandvik1999} when dealing with weakly coupled chains ($J_\perp/J\simeq 0.08$). For various system sizes, temperatures and Br-doping concentrations $x=10\%$, $12.5\%$ and $16.67\%$, we compute two different thermodynamic quantities, averaged over $300$ disorder samples for each point: the spin stiffness $\rho_s$~\cite{pollock1987,sandvik1997} and the transverse AF order parameter $m_x\equiv\sum_{i,j}\mathrm{e}^{i\mathbf{q}\cdot\mathbf{r}_{ij}}\langle S^+_iS^-_j\rangle/N^2$ at the AF wave-vector $\mathbf{q}=(\pi,\pi,\pi)$, which both reveal a finite temperature transition using a standard finite-size scaling analysis~\cite{sandvik2010},
\bea
    \rho_s(L)&=&L^{2-D}\,\, \mathcal{G}_{\rho_s}\left[L^{1/\nu}\left(T-T_c\right)\right]~\mathrm{and}\nonumber\\
    m_x(L)&=&L^{-\beta/\nu}\,\, \mathcal{G}_{m_x}\left[L^{1/\nu}\left(T-T_c\right)\right],
    \label{eq:stiff_scal_antz}
\eea
where $D=3$ is the dimensionality. The $3$D-XY critical exponents~\cite{burovski2006,campostrini2006,beach2005} $\nu=0.6717$ (the correlation length exponent) and $\beta=0.3486$ (the order parameter exponent) are used to extract the critical temperature $T_c$, after a Bayesian scaling analysis~\cite{harada2011,harada2015}. One can also include corrections to scaling of the form ${\mathcal{G}}\left[L^{1/\nu}\left(T-T_c\right)(1+cL^{-\omega})\right]$, where $\omega$ is a subleading exponent $[\mathcal{O}(1)]$ accounting for  a finite-size drift, which gives similar values within the error bars. Our final $T_c$ estimates (Fig.~\ref{fig:finite_temperature}) are averages of the individual $T_c$ from $\rho_s$ and $m_x$ crossings, with and without irrelevant corrections, while the given error bars reflect uncertainty between various estimates~\footnote{The critical temperature data for $x=10\%$ from $H=12.5\;\mathrm{T}$ to $H=13.0\;\mathrm{T}$ (Fig.~\ref{fig:finite_temperature}), are estimated using the $\beta$-doubling procedure, which strongly limits the temperature grid taken to perform the finite-size scaling analysis~\eqref{eq:stiff_scal_antz}: $1/T=2^p$ with $p\in\mathbb{N}$. Therefore, the error bars in this case reflect the closest available (min/max) temperatures to the estimated $T_c$ value.}.

\subsection{Finite temperature phase diagram}

\begin{figure}[t!]
    \includegraphics[width=1\columnwidth,clip,angle=0]{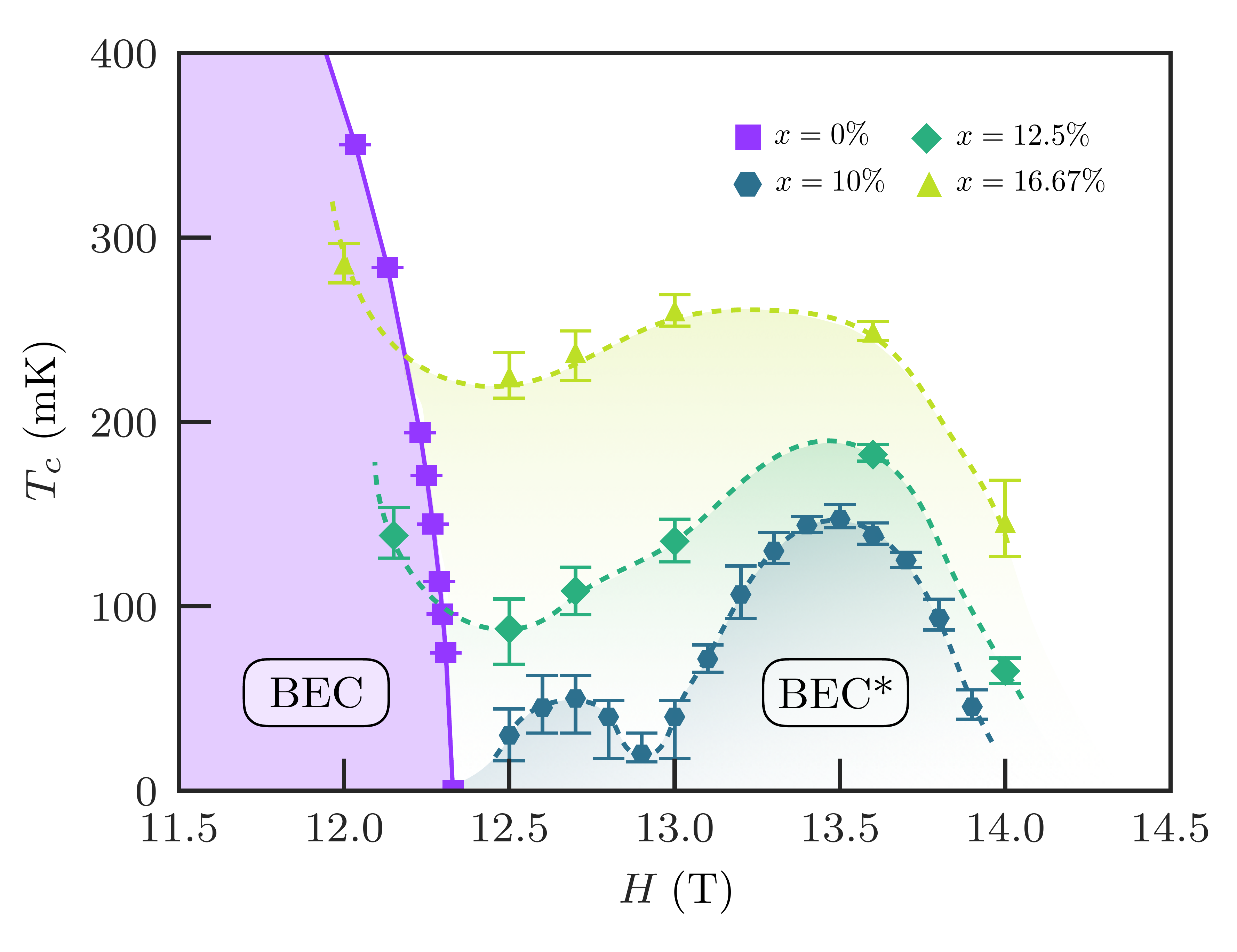}
    \caption{Finite temperature phase diagram at high magnetic field for various Br doping concentrations, obtained by QMC simulations of DTN$X$ [Eq.~\eqref{eq:DTNX}] for $x=0$ (square), $x=10\%$ (hexagon), $x=12.5\%$ (diamond), $x=16.67\%$ (triangle). Besides the clean BEC dome at $x=0$, a new impurity-induced ordered regime BEC$^*$ develops at higher magnetic field. While at large doping $x=12.5\%$ and $x=16.67\%$, BEC and BEC$^*$ overlap, for $x=10\%$ one clearly sees two resurgent  distinct BEC$^*$ mini-domes (see text).}
    \label{fig:finite_temperature}
\end{figure}

Fig.~\ref{fig:finite_temperature} shows the global magnetic field - temperature $H$--$T$ phase diagram obtained from extensive QMC simulations of the DTN$X$ model Eq.~\eqref{eq:DTNX} for various Br-impurity concentrations $x$. As previously discovered~\cite{dupont2017}, besides the clean BEC type order below $H_{c2}=12.3$ T, doping with Br leads to a new type of disorder-induced ordered phase, which we call BEC$^*$, appearing as a mini-dome centered around the single-impurity crossover field $H^*\simeq 13.6$ T.  This regime is quite extended and overlap with the clean BEC dome for $x> 10\%$. Interestingly, for $x=10\%$ a second mini-dome appears, centered around $12.7$ T and separated from the main BEC$^*$ phase made of single impurity states around $H^*$. This observation clearly confirms the expectation (see Section~\ref{sec:mie}) that objects made of two neighbouring impurities at distance $r_{\parallel}=2$, whose crossover field is precisley at $12.7$ T, should experience an effective interactions also leading to the long-range order.

The natural question opened by the observation of a second, ``satellite'' BEC$^*$ phase concerns the general trend when the impurity concentration gets more reduced: one can wonder whether more satellites may appear, and if intervening localized BG regimes could eventually be stabilized between these ordered phases. In order to address this fundamental issue, especially important to properly define the real extent of the high-field BG state proposed by Yu {\it{et al.}}~\cite{yu2012nature}, we now turn to GS physics at lower impurity concentration.

\section{Zero-temperature phase diagram and Bose-glass physics}\label{sec:bose_glass}

\begin{figure*}[t]
    \centering
    \includegraphics[width=2\columnwidth,clip,angle=0]{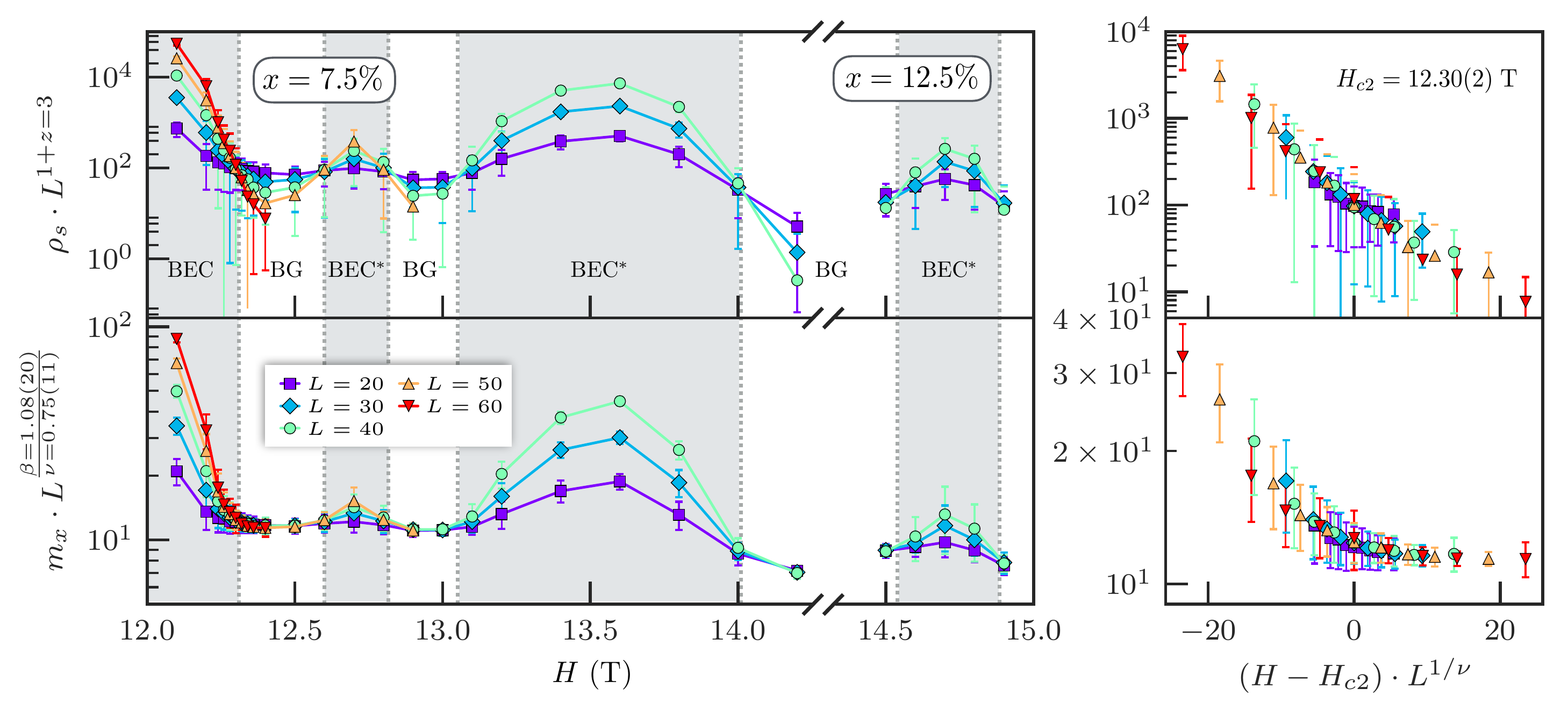}
    \caption{Left: finite size scaling analysis ($L=20,\ldots,60$) for the spin stiffness $\rho_s$ and the AF order parameter $m_x$ at zero temperature, from QMC simulations with the $\beta$-doubling scheme, for $x=7.5\%$ Br doping. For points above 14.3~T, the Br doping is taken to be $x=12.5\%$ to reduce numerical difficulties in computing $\rho_s$ and $m_x$ at high magnetization. Each point is  averaged over $200$ independant disordered samples. The dynamical exponent value was set exactly to $z=D=3$. The first QCP position $H_{c2}=\hcval\;\mathrm{T}$ and exponent $\nu=\nuexp$ were estimated from the $\rho_s$ data. Setting thus $H_{c2}$ and $\nu$, the order parameter exponent $\beta=\betaexp$ was then determined from the $m_x$ data. Right: scaling functions Eq.~\eqref{eq:stiff_scal_antz_H} for the data around $H_{c2}$, which gives quite good collapse, thus supporting that $z=D=3$.}
    \label{fig:bdoubling}
\end{figure*}

\subsection{Quantum Monte Carlo}\label{sec:qmc_bdoubling}

To study the $T=0$ phase diagram of DTN\textit{X} at high magnetic fields [Eq.~\eqref{eq:DTNX}], we use the QMC/SSE techniques again, but this time associated with the $\beta$-doubling scheme~\cite{sandvik2002} to reach low temperatures much faster than in standard schemes, in order to probe the GS properties. We remark that this method leads to large Monte Carlo errors, due to the purposely small number of performed thermalization and measurement steps, and may occasionally lead to systems out of the GS for some samples. Nevertheless, the estimate of the observables over different disorder realizations is reliable, as it gives larger statistical errors (sample-to-sample fluctuations) than the ones generated by the method. We compute the spin stiffness $\rho_s$ and the transverse order parameter $m_x$, averaged over $200$ different samples for each of the points presented in Figure~\ref{fig:bdoubling}. The finite size scaling analysis close to the BEC--BG transition follows
\bea
    \rho_s(L)&=&L^{2-D-z}\,\, \mathcal{G}_{\rho_s}\left(L^{1/\nu}\left|H-H_c\right|\right)~\mathrm{and}\nonumber\\
    m_x(L)&=&L^{-\beta/\nu}\,\, \mathcal{G}_{m_x}\left(L^{1/\nu}\left|H-H_c\right|\right),
    \label{eq:stiff_scal_antz_H}
\eea
where $H_c$ is the critical field, $D=3$ is the dimensionality and $z$ is the dynamical exponent.

\subsection{Zero temperature phase diagram}

The zero-temperature phase diagram of DTN$X$ at high magnetic field is shown in Fig.~\ref{fig:bdoubling} for the Br doping concentration $x=7.5\%$.
Both the main BEC$^*$ phase and its left ``satellite'' centered around $12.7$ T get reduced as compared to their $x=10\%$ extension, and two intervening localized BG phases are stabilized in between. Above 14 T, a third BG regime is also observed. Interestingly, ordered phase appears at each of the noticeable levels crossings $H=12.7\;\mathrm{T}$, $H=13.6\;\mathrm{T}$ and $H=14.7\;\mathrm{T}$ as previously anticipated in Section~\ref{sec:mie}. For the highest field value, this feature is also reported, but  for a higher doping concentration ($x=12.5\%$), because of numerical difficulties to capture GS properties close to saturation (which corresponds to very small density of particles).

\subsection{The BEC-BG phase transition}
The quantum phase transition between BEC and BG phases remains controversial in various aspects, such as the precise value of some critical exponents~\cite{priyadarshee2006,weichman2007,gurarie2009,meier2012,alvarez2015,ng2015}. While it is now well established that the correlation length exponent $\nu$ satisfies the Harris-Chayes bound $\nu\ge 2/D$~\cite{harris1974,chayes1986}, there are still some debates regarding the dynamical exponent equality $z=D$~\cite{priyadarshee2006,weichman2007,meier2012,alvarez2015,ng2015}, as well as for the exponent $\phi$ governing the critical temperature $T_c\sim |H-H_c|^{\phi}$, for which some recent results~\cite{yu2012nature,yu2012} are inconsistent with the theoretical bound $\phi\ge 2$~\cite{fisher1989}, verified in a more recent numerical study~\cite{yao2014}.
Besides these theoretical discussions, only a few experimental realizations of dirty bosons are available to test such predictions, and in particular for condensed matter systems.  In Tl$_x$K$_{(1-x)}$CuCl$_3$~\cite{yamada2011} and (C$_4$H$_{12}$N$_2$)Cu$_2$(Cl$_{(1-x)}$Br$_x$)$_6$~\cite{huvonen2012}, as well as in DTN$X$~\cite{yu2012nature}, the measured exponent of the critical boundary was found to be $\phi\sim 1$, which lies clearly below the $\phi=2$ bound.

In the following we address some critical properties of the $T=0$ BEC-BG transition close to $H_{c2}$, the critical field ending the ordered phase of the clean degrees of freedom, for a doping level $x=7.5\%$, as shown in Fig.~\ref{fig:bdoubling}. We start with the finite size scaling analysis of the spin stiffness $\rho_s$, setting the dynamical exponent to exactly $z=3$. This leads to a very nice single-point crossing for the different system sizes $L=20,30,40,50$ and $60$, meaning that there is a QPT happening at the crossing point, $H_{c2}=\hcval\;\mathrm{T}$. This value for the QCP is identical to the value of $H_{c2}^\mathrm{clean}$ in pure DTN, suggesting that the degrees of freedom defining the end of this first ordered phase are the ``clean'' spins. By optimizing the collapse of the data sets obtained for different $L$ values on a single (scaling) curve, one can estimate the correlation length exponent $\nu=\nuexp>2/3$, compatible with the Harris-Chayes criterion. Using these estimates of $H_{c2}$ and $\nu$, we perform a similar finite size scaling analysis for the AF order parameter $m_x$, and get the exponent $\beta=\betaexp$, in agreement with the previous work~\cite{yu2012}.  The scaling collapse of $m_x$ data is also very good, confirming the value of $\nu$ obtained from  $\rho_s$ data. Through the hyperscaling relation, the anomalous exponent $\eta$ is found to be
\be
    \eta = 2\beta/\nu-D-z+2=\etaexp,
\ee
which verifies the inequality $\eta\leq 2-D=-1$~\cite{fisher1989}.  Overall, the dynamical exponent value $z=D=3$ is fully compatible with our results, confirming previous studies~\cite{hitchcock2006,yu2012,yao2014}. Note also that good crossings are obtained at the other BG-BEC$^*$ transitions, as visible in Fig.~\ref{fig:bdoubling}.

We have not directly addressed the so-called ``$\phi$-crisis'' raised by conflicting numerics~\cite{yao2014,yu2014}. It is clearly a very difficult numerical task to safely probe the quantum critical regime using finite temperature data. Moreover, we believe that the very peculiar situation at play in DTN$X$, with successive narrow BEC$^*$ and BG regimes, is not favorable to disentangle a genuine quantum critical regime from crossover effects due to competing phases.

\section{Conclusions}
\label{sec:conclusion}

In a first step, based on recent NMR experiments at high magnetic field~\cite{orlova2017} we have fully determined in Sec.~\ref{sec:theo_model} the microscopic model of the DTN$X$ compound. Indeed, these experimental results can be interpreted and understood via single impurity physics, which makes it possible to perform analytical as well as exact diagonalization calculations on large systems from which a unique set of coupling parameters~\eqref{eq:DTNX_parameters} can be determined for the impurity degrees of freedom. Moreover, this simple description provides fruitful insights on the picture of DTN$X$ at high magnetic field, such as the strong localization of isolated impurity states and the fact that the clean background polarizes for a smaller magnetic field than the impurities. Thus, a simple picture of DTN$X$ at high magnetic field consists in a frozen (clean) background with a collection of impurities spatially randomly distributed, yet to be polarized upon increasing the magnetic field.

A natural extension was then to study the mutual effect of two impurities, which was done in Sec.~\ref{sec:manybody_impurity}. By means of ED, we reveal that, despite the strong localization of the impurity states, there exists an effective unfrustrated pairwise interaction between impurity degrees of freedom. In order to capture the relevant low-energy physics we have built an effective model of bosons in a diluted lattice with an exponentially decaying coupling with the distance between bosons. This model suggests that the bosonic degrees of freedom can order at low-enough temperature, which is confirmed by recent QMC simulations of the full microscopic model in Ref.~\onlinecite{dupont2017}. This paves the way to a resurgence of global phase coherence in DTN$X$, in sharp contrast with the uninterrupted many-body localized Bose-glass phase reported in Ref.~\onlinecite{yu2012nature}.

In Sec.~\ref{sec:DTNX_physics} we have extended the finite-temperature study of the realistic DTN$X$ Hamiltonian with state-of-the-art QMC simulations at lower temperature, for a Br concentration $x=10\%$, in order to compute the extension of the disorder-induced BEC$^*$ revival and of the BG regime. We have first shown that, for this concentration, the BEC$^*$ is connected to the large BEC phase of the clean sites without any intervening BG. Furthermore, we reveal that the critical temperature boundary of the BEC$^*$ actually presents not one, but two distinct domes: the expected one centered at $H^*\sim 13.6\;\mathrm{T}$, which corresponds to the condensation of single impurity degrees of freedom, and a new one centered around $H\sim 12.7\;\mathrm{T}$. The new dome can be understood as the ordering of multi-impurity objects, thoroughly discussed in Sec.~\ref{sec:mie}. This considerably extends the current picture of the phase diagram of DTN$X$ at high magnetic field: at low enough doping concentration, the consecutive disorder-induced BEC$^*$ mini-domes are separated by intervening many-body localized BG regimes . However, decreasing the doping concentration makes it very hard to reliably obtain the critical temperature in numerical simulations. Consequently, in Sec.~\ref{sec:bose_glass} we rather turned our attention to $T=0$ physics for $x=7.5\%$, focusing on the still controversial quantum phase transition between the BEC and BG phases, in order to determine the critical exponents.

Finally, we now expect experimental investigations of the suggested disorder-induced BEC$^*$ at high magnetic field in DTN$X$ to confirm our theoretical results. The single-impurity BEC$^*$ dome centered around $H^*\sim 13.6\;\mathrm{T}$ should be easily accessible to experiments, as the estimated critical temperatures are above $100$ mK for higher doping levels. Similarly, for $x\sim 10\%$ one should be able to probe the upper part of the $H\sim 12.7\;\mathrm{T}$ dome with estimated $T_c$ around $50\;\mathrm{mK}$. This second observation is experimentally very challenging, but it would definitely confirm the overall understanding of the high-magnetic field phase diagram of DTN$X$ presented in this paper, consisting of alternated ordered and many-body localized phases. While DTN$X$ was previously proposed as an excellent model material to study the BG phase and the  BG-BEC phase transitions, we have found that in DTN$X$ the genuine properties of the QPT may be spoiled by closely surrounding disorder-induced LRO phases. With this respect, some of the BEC$^*$-BG phase boundaries may be ``better'' than the others, but in all these cases, the required temperature range representative of the critical behavior appears to be prohibitively low.

\begin{acknowledgments}
We would like to thank Tommaso Roscilde for discussions. NL acknowledges Markus M\"uller for interesting discussions regarding the Lifshitz model~\cite{lifshitz1965}.
This work was performed using HPC resources from GENCI (Grant No. x2016050225 and No. x2017050225) and CALMIP. We acknowledge support of the French ANR program BOLODISS (Grant No. ANR-14-CE32-0018), R\'egion Midi-Pyr\'en\'ees and the Condensed Matter Theory Visitors Program at Boston University, and Programme Investissements d'Avenir within the ANR-11-IDEX-0002-02 program, reference ANR-10-LABX-0037-NEXT.
\end{acknowledgments}

\end{document}